\documentclass[11pt,a4paper]{article}
\pdfoutput=1
\usepackage{putex}
\usepackage[T1]{fontenc}
\usepackage[utf8]{inputenc}
\usepackage{amsmath,amssymb,amsthm,mathrsfs}
\usepackage{bbm}
\usepackage{microtype}
\usepackage{cite}
\usepackage{xcolor}
\usepackage{booktabs}
\usepackage{array}
\usepackage{tikz}
\usetikzlibrary{arrows.meta}
\usepackage{hyperref}
\hypersetup{colorlinks=true,linkcolor=blue!60!black,citecolor=green!40!black,urlcolor=blue!60!black}

\newcommand{\RR}{\mathbb{R}}
\newcommand{\CC}{\mathbb{C}}
\newcommand{\QQ}{\mathbb{Q}}
\newcommand{\ZZ}{\mathbb{Z}}
\newcommand{\NN}{\mathbb{N}}
\newcommand{\HH}{\mathbb{H}}
\newcommand{\Ds}{\mathscr{D}}
\newcommand{\cH}{\mathcal{H}}
\newcommand{\cM}{\mathcal{M}}
\newcommand{\cO}{\mathcal{O}}
\newcommand{\cV}{\mathcal{V}}
\newcommand{\cZ}{\mathcal{Z}}
\newcommand{\cE}{\mathcal{E}}
\newcommand{\Ginf}{\Gamma_{\infty}}
\newcommand{\Mcore}{C}
\newcommand{\e}{\mathbf{e}}
\newcommand{\btau}{\boldsymbol{\tau}}
\newcommand{\bgamma}{\boldsymbol{\gamma}}
\newcommand{\bx}{\boldsymbol{x}}
\newcommand{\bdelta}{\boldsymbol{\delta}}
\newcommand{\bk}{\boldsymbol{k}}
\newcommand{\ket}[1]{|#1\rangle}
\newcommand{\bra}[1]{\langle #1|}
\newcommand{\vev}[1]{\left\langle #1\right\rangle}
\newcommand{\doteqf}{\mathrel{\overset{\scriptscriptstyle\mathrm{fr}}{=}}}
\DeclareMathOperator{\SL}{SL}
\DeclareMathOperator{\Sp}{Sp}
\DeclareMathOperator{\Mp}{Mp}
\DeclareMathOperator{\GL}{GL}
\DeclareMathOperator{\Aut}{Aut}

\numberwithin{equation}{section}
\makeatletter
\g@addto@macro\normalsize{\abovedisplayskip=11pt plus 3pt\belowdisplayskip=11pt plus 3pt\abovedisplayshortskip=\abovedisplayskip\belowdisplayshortskip=\belowdisplayskip}
\makeatother
\AtBeginDocument{\normalsize}
\newtheorem{proposition}{Proposition}[section]

\preprint{RIKEN-iTHEMS-Report-26}
\title{Summing over Seifert Manifolds \\ in Narain Ensemble Averages}
\authors{\Large Kei Kaneda$^{1}$\footnote{\href{mailto:kaneda@hep-th.phys.s.u-tokyo.ac.jp}{kaneda@hep-th.phys.s.u-tokyo.ac.jp}} and Masahito Yamazaki$^{1,2,3,4}$\footnote{\href{mailto:yamazaki@hep-th.phys.s.u-tokyo.ac.jp}{yamazaki@hep-th.phys.s.u-tokyo.ac.jp}}}
\date{}
\institution{}{\centerline{$^{1}$Department of Physics, University of Tokyo, Tokyo 113-0033, Japan}}
\institution{}{\centerline{$^{2}$Kavli Institute for Physics and Mathematics of the Universe (Kavli IPMU),}}
\institution{}{\centerline{UTIAS, University of Tokyo, Chiba 277-8583, Japan}}
\institution{}{\centerline{$^{3}$Trans-Scale Quantum Science Institute, University of Tokyo, Tokyo 113-0033, Japan}}
\institution{}{\centerline{$^{4}$Center for Interdisciplinary Theoretical and Mathematical Sciences (iTHEMS),}}
\institution{}{\centerline{RIKEN, Saitama 351-0198, Japan}}

\begin{document}

\abstract{The ensemble average of Narain conformal field theories over their moduli space is a real-analytic Eisenstein series, whose Poincar\'e expansion can be interpreted as a sum over lens spaces weighted by the partition functions of Abelian Chern--Simons theory. In this paper we extend this correspondence from lens spaces to Seifert manifolds, which gives an example of a restricted sum over topologies of the kind that arises in three-dimensional quantum gravity. A Seifert manifold is obtained from a fixed core, a punctured surface times a circle, by gluing solid tori to its boundary tori, and we place a copy of the Narain theory on each of these tori. 
The ensemble average supplies the sum over the solid tori, or more precisely over marked presentations with fixed base genus, number of fibers and integer Seifert invariant. The resulting sum converges absolutely and is a polynomial in the components of the vector-valued Eisenstein series. 
When all the boundary tori share a single modulus, the average acquires connected corrections, which we identify with wormholes gluing the core to itself; for two boundary tori, the terms with rational Lagrangian sublattices of the Siegel--Eisenstein series are realized by wormholes containing an exceptional fiber. After reduction along the fiber, the construction closely parallels the two-dimensional topological gravity of Marolf and Maxfield, with the base of the fibration playing the role of their spacetime. We also discuss Dehn fillings of 3-manifolds with torus boundary, non-orientable bases, and the surgery formula. Our results suggest that modularity could be an organizing principle in the summation of geometries in quantum gravity.
}

\pagenumbering{Alph} 
\maketitle
\pagenumbering{arabic}
\tableofcontents

\section{Introduction and Summary}
\label{sec:intro}

The role of topology in quantum gravity has been debated for decades: should the gravitational path integral include a sum over spacetime topologies, and if so, which topologies contribute and with what weights~\cite{Coleman:1988cy,Giddings:1988wv,Banks:1988je,Banks:1989zw,Fischler:1989ka}? In three dimensions the variety of 3-manifolds makes a complete sum hard to define, let alone to control~\cite{Carlip:1992us,Carlip:1992wg,Banerjee:2022pmw}. Much of the recent work has therefore focused on sums over restricted classes of geometries, such as the handlebodies that contribute to holographic partition functions~\cite{Dijkgraaf:2000fq,Maloney:2007ud,Maloney:2020nni,Afkhami-Jeddi:2020ezh}, on sums over bordisms in topological field theory~\cite{Marolf:2020xie,Banerjee:2022pmw}, and more recently on sums over topologies in ``TQFT gravity'' and in pure three-dimensional gravity~\cite{Barbar:2023ncl,Dymarsky:2024frx,Dymarsky:2025agh,Angelinos:2025zek,Nicosanti:2025xwu,Belin:2026pko,Dymarsky:2026asf,Dymarsky:2026lnf}. In two dimensions, the relation between such sums and ensemble averages is much better understood~\cite{Saad:2019lba,Marolf:2020xie}. In this paper we work out an example of a sum over geometries in three-dimensional Abelian Chern--Simons theory: the ensemble average of Narain CFTs generates a sum over marked presentations of all Seifert manifolds, with explicitly computable weights.

Our starting point is as follows. Let $\Lambda\cong\ZZ^{p+q}$ be a lattice with an even integral quadratic form $Q$ with $p$ positive and $q$ negative eigenvalues, and let $\cM_Q$ be the corresponding Narain moduli space~\cite{Narain:1985jj,Narain:1986am}, with the discriminant group marked (Section~\ref{sec:review}). For $p+q>4$ and $pq\neq0$, the Siegel--Weil formula states that the Narain average of the Siegel--Narain theta function $\Theta_{x}(\tau;m)$, $x\in\Ds=\Lambda^*/\Lambda$, is a vector-valued Eisenstein series~\cite{Siegel1951,Weil1965,Ashwinkumar:2021kav,Maloney:2020nni,Afkhami-Jeddi:2020ezh,Benjamin:2021wzr}:
\begin{equation}
  \vev{\Theta_x(\tau;m)}_{\cM_Q}=E_x(\tau)
  =\sum_{\gamma\in\Ginf\backslash\SL(2,\ZZ)}
  \frac{\bra{x}\rho(\gamma)^{-1}\ket{0}}{(c\tau+d)^{p/2}(c\bar\tau+d)^{q/2}}\;.
  \label{eq:intro-adkly}
\end{equation}
Here $\tau$ is the modulus of the torus on which the CFT lives and $m\in\cM_Q$ is the Narain modulus. The functions $\Theta_x$, defined in~\eqref{eq:theta-def}, are the components of a vector $\Theta(\tau;m)=\sum_x\Theta_x(\tau;m)\ket{x}\in\CC[\Ds]$, on which $\SL(2,\ZZ)$ acts by the Weil representation $\rho$ of~\eqref{eq:weil}. The sum runs over cosets of $\Ginf=\{\pm T^k\mid k\in\ZZ\}$, and $(c,d)$ is the lower row of $\gamma$.

In~\cite{Ashwinkumar:2021kav} the coefficient $\bra{x}\rho(\gamma)^{-1}\ket{0}$ was identified with the partition function of the $U(1)^{p+q}$ Chern--Simons theory with kinetic matrix $Q$ on the lens space $L(c,d)$ (up to orientation conventions), decorated by a Wilson line $x$. The average over Narain CFTs can thus be rewritten as a sum over lens spaces. This setup has since been extended to orbifolds of generalized Narain theories~\cite{Ashwinkumar:2023eit}, and the averaged theory has been found to exhibit emergent symmetries which are absent from its individual members~\cite{Ashwinkumar:2023jtz,Kawamoto:2026jdz,Yamazaki:2026ppe}.

The Seifert fibered spaces over $S^2$ with at most two exceptional fibers are precisely the lens spaces, including $S^3$ and $S^2\times S^1$. It is then natural to ask whether one can sum over \emph{all} Seifert manifolds $M(g;\chi;(\alpha_1,\beta_1),\dots,(\alpha_n,\beta_n))$ and still obtain a modular object which is realized by an ensemble average. Here $g$ is the genus of the base, the coprime pairs $(\alpha_i,\beta_i)$ specify the fibers, and $\chi\in\ZZ$ is an integer contribution to the Euler number~\eqref{eq:euler}; the precise definition is~\eqref{eq:seifert-def}.

Seifert manifolds are the next class after lens spaces, since they remain tractable under circle reduction and their Chern--Simons invariants are known in closed form~\cite{Freed:1991wd,Rozansky:1993zx,LawrenceRozansky1999,Blau:2013oha}. Their exceptional fibers are the topological counterpart of conical defects, which have been proposed as additional contributions to the gravitational path integral: orbifolds were included as saddle points in~\cite{Benjamin:2020mfz}, and Seifert manifolds, which do not admit a classical solution, in~\cite{Maxfield:2020ale}. In pure gravity, the sum over genus-zero Seifert manifolds has been reduced by Dehn surgery to the partition function of $\Sigma_{0,n}\times S^1$~\cite{deBoer:2025oge}, which has a similar structure to our decomposition~\eqref{eq:intro-decomp}. Seifert manifolds whose base has genus $g\ge1$ or several boundary components naturally involve several CFT surfaces at once, as in the recent discussions of wormholes in ensemble holography~\cite{Saad:2019lba,Maloney:2020nni,Cotler:2020ugk,Collier:2021rsn,Kawamoto:2026jdz}. Geometries of the form $\Sigma\times S^1$, most of which do not admit a classical solution, have also been studied in three-dimensional chiral gravity~\cite{Eberhardt:2022wlc}. Finally, it has been asked whether a sum over bordisms in three dimensions becomes more manageable when restricted to a suitable class of 3-manifolds~\cite{Banerjee:2022pmw}, and our construction gives one such restricted sum, although its class is not closed under gluing.

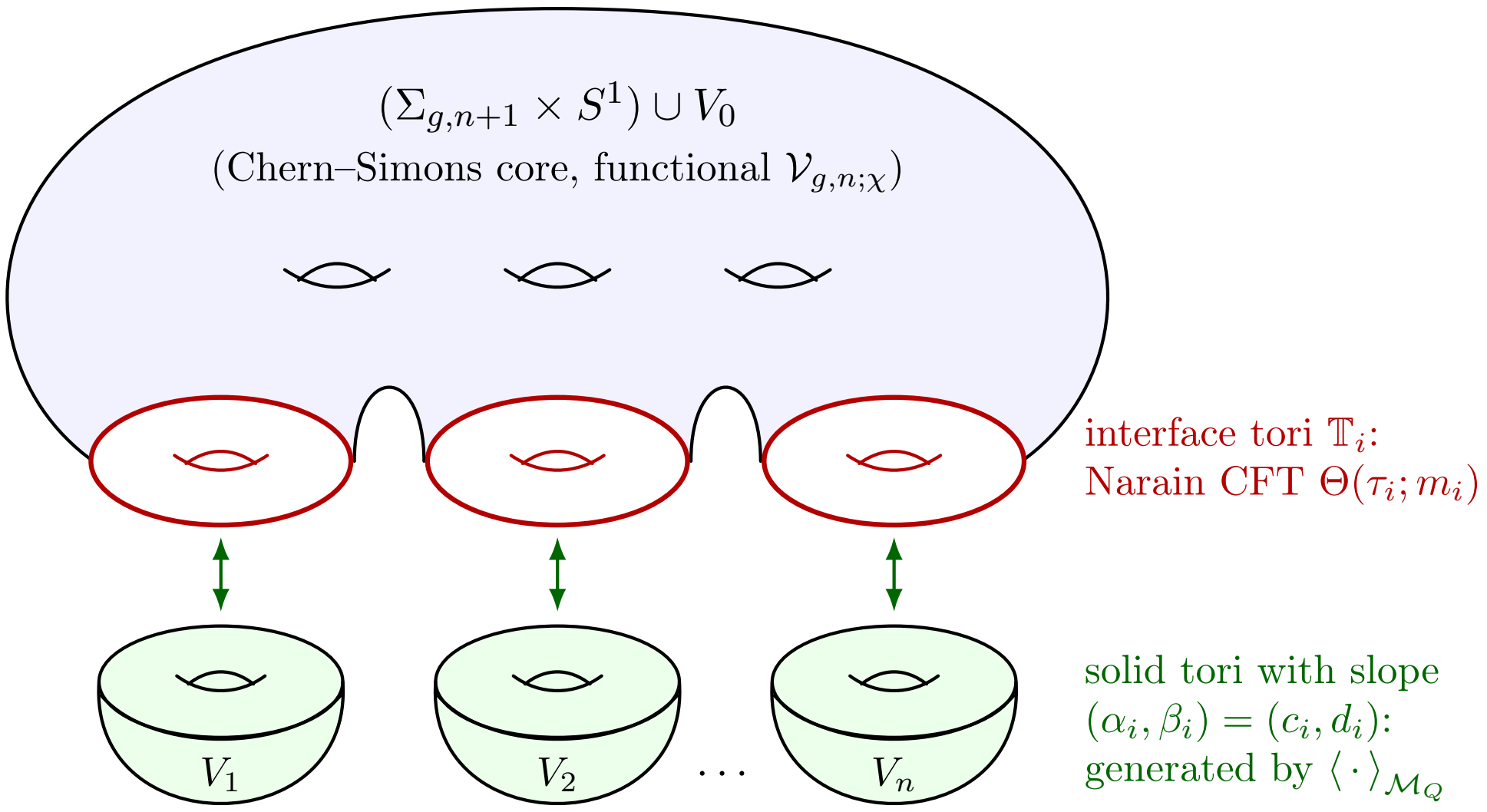
\begin{figure}[t]
  \centering
  \begin{tikzpicture}[line cap=round,line join=round,thick]
    \fill[blue!5] (-4.0,-0.6) .. controls (-5.4,0.4) and (-5.2,3.3) .. (0,3.3)
    .. controls (5.2,3.3) and (5.4,0.4) .. (4.0,-0.6) -- (1.75,-0.6)
    .. controls (1.75,0.25) and (1.15,0.25) .. (1.15,-0.6) -- (-1.15,-0.6)
    .. controls (-1.15,0.25) and (-1.75,0.25) .. (-1.75,-0.6) -- cycle;
    \draw (-4.0,-0.6) .. controls (-5.4,0.4) and (-5.2,3.3) .. (0,3.3)
    .. controls (5.2,3.3) and (5.4,0.4) .. (4.0,-0.6);
    \draw (1.75,-0.6) .. controls (1.75,0.25) and (1.15,0.25) .. (1.15,-0.6);
    \draw (-1.15,-0.6) .. controls (-1.15,0.25) and (-1.75,0.25) .. (-1.75,-0.6);
    \foreach \x in {-1.9,0,1.9}{
        \draw (\x-0.45,1.05) .. controls (\x-0.15,0.85) and (\x+0.15,0.85) .. (\x+0.45,1.05);
        \draw (\x-0.33,0.96) .. controls (\x-0.1,1.15) and (\x+0.1,1.15) .. (\x+0.33,0.96);
      }
    \node at (0,2.45) {$(\Sigma_{g,n+1}\times S^1)\cup V_0$};
    \node[font=\small] at (0,1.9) {(Chern--Simons core, functional $\cV_{g,n;\chi}$)};
    \foreach \x/\lab in {-2.9/1, 0/2, 2.9/n} {
        \draw[very thick,red!70!black,fill=white] (\x,-0.6) ellipse (1.12 and 0.55);
        \draw[red!70!black] (\x-0.4,-0.55) .. controls (\x-0.15,-0.72) and (\x+0.15,-0.72) .. (\x+0.4,-0.55);
        \draw[red!70!black] (\x-0.3,-0.62) .. controls (\x-0.1,-0.48) and (\x+0.1,-0.48) .. (\x+0.3,-0.62);
        \draw[{Latex[length=2mm]}-{Latex[length=2mm]},green!40!black] (\x,-1.25) -- (\x,-1.9);
        \fill[green!8] (\x-1.05,-2.5) .. controls (\x-1.15,-3.9) and (\x+1.15,-3.9) .. (\x+1.05,-2.5) -- cycle;
        \fill[green!8] (\x,-2.5) ellipse (1.05 and 0.48);
        \draw (\x,-2.5) ellipse (1.05 and 0.48);
        \draw (\x-1.05,-2.5) .. controls (\x-1.15,-3.9) and (\x+1.15,-3.9) .. (\x+1.05,-2.5);
        \draw (\x-1.03,-2.62) .. controls (\x-0.6,-3.12) and (\x+0.6,-3.12) .. (\x+1.03,-2.62);
        \draw (\x-0.38,-2.45) .. controls (\x-0.15,-2.62) and (\x+0.15,-2.62) .. (\x+0.38,-2.45);
        \draw (\x-0.28,-2.52) .. controls (\x-0.1,-2.38) and (\x+0.1,-2.38) .. (\x+0.28,-2.52);
        \node[font=\small] at (\x,-3.3) {$V_{\lab}$};
      }
    \node at (1.45,-3.3) {$\cdots$};
    \node[red!70!black,font=\small,align=left,anchor=west] at (4.4,-0.6) {interface tori $\mathbb T_i$:\\ Narain CFT $\Theta(\tau_i;m_i)$};
    \node[green!40!black,font=\small,align=left,anchor=west] at (4.4,-2.9) {solid tori with slope\\ $(\alpha_i,\beta_i)=(c_i,d_i)$:\\ generated by $\vev{\,\cdot\,}_{\cM_Q}$};
  \end{tikzpicture}
  \caption{The Narain theory lives on the tori that separate the core from tubular neighborhoods of the exceptional fibers. Averaging over the Narain moduli fills each interface with a sum over solid tori, which produces a sum over Seifert manifolds. The lens-space sum of~\cite{Ashwinkumar:2021kav} is the case of a single interface whose other side is a solid torus. The solid torus $V_0$ is part of the core and is not drawn separately.}
  \label{fig:picture}
\end{figure}

A Seifert manifold decomposes as
\begin{equation}
  M=\underbrace{\bigl(\Sigma_{g,n+1}\times S^1\bigr)\cup V_0}_{\text{core (fixed)}}\ \cup\ \underbrace{V_1\sqcup\dots\sqcup V_n}_{\text{solid tori (summed)}}\;.
  \label{eq:intro-decomp}
\end{equation}
Here $\Sigma_{g,n+1}$ is a genus-$g$ surface with $n+1$ boundary components, the solid torus $V_0$ is glued along the slope $(1,\chi)$, and each $V_i$ with $i\ge1$ is glued along the slope $(\alpha_i,\beta_i)$ (see~\eqref{eq:seifert-def}). We place a copy of the Narain theory on each of the $n$ interface tori. The Chern--Simons theory on the core supplies a finite-dimensional functional $\cV_{g,n;\chi}$ on the interface Hilbert spaces $\CC[\Ds]^{\otimes n}$, and the ensemble average supplies the sum over the solid tori $V_i$, as in~\eqref{eq:intro-adkly}. The setup is shown in Figure~\ref{fig:picture}.

In the basis labeled by the holonomy around the fiber, the Chern--Simons functional of the core turns out to be diagonal,
\begin{equation}
  \cV_{g,n;\chi}=|\Ds|^{g-1+n/2}\sum_x\e(-\chi q(x))\,\bra{x}^{\otimes n}\;,
  \label{eq:intro-core}
\end{equation}
where $q(x)=\tfrac12Q(x,x)\bmod1$ is the quadratic form on $\Ds$~\eqref{eq:disc-form-pairing}, $\bra{x}^{\otimes n}:=\bra{x}_1\otimes\cdots\otimes\bra{x}_n$, and $\e(z):=e^{2\pi i z}$. Hence the Abelian Chern--Simons partition function of any Seifert manifold is a finite sum of products of the lens-space Gauss sums $\varepsilon_{Q,x}(\alpha_i,\beta_i)$ of~\cite{Ashwinkumar:2021kav} (Section~\ref{sec:seifert}), which are the coefficients in~\eqref{eq:intro-adkly} with $(c,d)=(\alpha_i,\beta_i)$ (see~\eqref{eq:eps-def} and~\eqref{eq:eps}).

Contracting $\cV_{g,n;\chi}$ with the solid-torus states $\rho(\gamma_i)^{-1}\ket0$ defines, for each marked presentation $(g;\chi;\gamma_1,\dots,\gamma_n)$ with $\gamma_i\in\Ginf\backslash\SL(2,\ZZ)$ of lower row $(\alpha_i,\beta_i)$, an amplitude $Z^{\rm pr}_{\rm CS}[g;\chi;\gamma_1,\dots,\gamma_n]$ as in~\eqref{eq:ZM}. It is the Chern--Simons partition function of the Seifert manifold $M(g;\chi;\gamma_1,\dots,\gamma_n)$ in the framing fixed by the presentation, and it differs from the canonically framed invariant by an explicit power of $\e(\sigma/8)$, where $\sigma=p-q$ is the signature (see~\eqref{eq:ZMcan}).

When the moduli $m_1,\dots,m_n$ of the $n$ Narain theories are independent and $p+q$ is even, the sum over \emph{marked presentations} with fixed $(g,n,\chi)$, which we call the \emph{Seifert--Eisenstein series}, is given by
\begin{equation}
  \begin{aligned}
      \sum_{\gamma_1,\dots,\gamma_n\in\Ginf\backslash\SL(2,\ZZ)}
      \frac{Z^{\rm pr}_{\rm CS}\bigl[g;\chi;\gamma_1,\ldots,\gamma_n\bigr]}{\prod_i(\alpha_i\tau_i+\beta_i)^{p/2}(\alpha_i\bar\tau_i+\beta_i)^{q/2}}
       & =|\Ds|^{g-1+\frac n2}\sum_{x\in\Ds}\e(-\chi q(x))\prod_{i=1}^nE_x(\tau_i)                                    \\
       & =\Bigl\langle \cV_{g,n;\chi}\bigl(\Theta(\tau_1;m_1)\otimes\cdots\otimes\Theta(\tau_n;m_n)\bigr)\Bigr\rangle
    \end{aligned}
  \label{eq:intro-main}
\end{equation}
(Proposition~\ref{prop:main}). Here $\tau_i$ and $m_i$ are the moduli of the $i$-th torus and Narain theory, and the bracket is the average of each $m_i$ over $\cM_Q$. Only the degree-one Siegel--Weil formula is needed here. The sum runs over marked presentations rather than over homeomorphism classes, so each Seifert manifold appears many times, related by Seifert moves~\eqref{eq:seifert-move} and relabelings of the fibers. The sum necessarily contains the identity-coset terms $\delta_{x,0}$ of the $E_x$, i.e.\ the terms with $(c,d)=(0,1)$ in~\eqref{eq:intro-adkly}, which correspond to fillings in which the fiber bounds a disc. These terms are in general connected sums of lens spaces and copies of $S^2\times S^1$, which need not be Seifert manifolds, but they are required by modular invariance.

The Seifert--Eisenstein series is a component of a finite-dimensional real-analytic vector-valued modular form for $\SL(2,\ZZ)^n$, namely the multiplet obtained by contracting $E(\tau_1)\otimes\cdots\otimes E(\tau_n)$ with arbitrary functionals. In particular, it transforms as a scalar modular form under the congruence subgroup $\Gamma(N)^n$, where $N$ is the level of $(\Ds,q)$. The modular $T$-transformation acting on $\tau_i$ shifts $\chi$, which is the Seifert move. The $S$-transformation exchanges the roles of the section and the fiber on the $i$-th torus, and maps $\cV_{g,n;\chi}$ to a functional that is no longer diagonal, so the Seifert sums alone do not close under the modular group. The same argument applies to an arbitrary 3-manifold with torus boundary, and gives what we call a Dehn-filling Eisenstein series.

When all the tori share a single modulus $m$, the average is instead given by the degree-$n$ Siegel--Eisenstein series contracted with $\cV_{g,n;\chi}$ (Section~\ref{sec:common}). Its decomposition by indecomposable Lagrangian sublattices coincides with the cumulant expansion, so that the average is the Seifert sum~\eqref{eq:intro-main} plus connected terms, and for $n=2$ we find that the connected terms glue the core to itself. More precisely, for each indecomposable Lagrangian we construct a representative cobordism that realizes its Chern--Simons state. For integral Lagrangians these are the torus wormholes of~\cite{Maloney:2020nni,Cotler:2020ugk,Collier:2021rsn}, while for rational ones they are wormholes containing one exceptional fiber, whose order is the denominator of the Lagrangian. The Lagrangian determines the state of the cobordism, while its topology is left undetermined. Two notions of connectedness are thus distinguished: the tori are topologically connected through the core even for independent moduli, while the common modulus makes them statistically connected, which gives the wormhole terms.

Finally, we reduce the construction along the fiber (Section~\ref{sec:2dgrav}). Abelian Chern--Simons theory then becomes the two-dimensional TQFT of the commutative Frobenius algebra $\mathrm{Fun}(\Ds)$ of functions on $\Ds$, and the Seifert sum is a genus-$g$ correlator of ``Eisenstein insertions'' $\cE(\tau)=\sum_xE_x(\tau)\tilde e_x$, where $\tilde e_x$ are the normalized delta functions on $\Ds$ (Section~\ref{sec:2d}). In this form it closely parallels the topological gravity of Marolf and Maxfield~\cite{Marolf:2020xie}. The base of the fibration plays the role of their two-dimensional spacetime, there are $|\Ds|$ sectors, the coupling $e^{S_0}$ is fixed to $|\Ds|^{-1/2}$, and their boundary observable $Z$ is replaced by $|\Ds|^{1/2}E_x(\tau)$. The sum over the genus of the base converges only if one introduces an additional weight, and pure Chern--Simons theory lies outside the region of convergence. With this weight, and for unimodular lattices, the formal generating functions of the two models coincide.

We treat several further topics more briefly. Appendix~\ref{app:surgery} gives the precise relation, including orientation and framing phase, between our formula for $Z_{\rm CS}$ and the surgery formula of~\cite{MOO1992,Deloup1999}. Section~\ref{sec:nonorientable} discusses non-orientable bases, for which the crosscaps project onto the $2$-torsion of $\Ds$ weighted by Frobenius--Schur indicators. Appendix~\ref{sec:orbifolds} discusses the $\ZZ_m$ orbifolds excluded in~\cite{Ashwinkumar:2021kav}, which differ from the lens-space terms only in their gravitational weight. An explicit example with $\Ds=\ZZ_3$ is worked out in Section~\ref{sec:example}, and we conclude in Section~\ref{sec:discussion} with a discussion of open problems.

Once the core functional is known, most of the derivations below are elementary; the main point is to identify the ensemble observable that produces each topological sum and to see why the result is modular.
\section{Narain averages and the topological dictionary}
\label{sec:review}
In this section we collect the facts about Narain averages and Abelian Chern--Simons theory that we need, mostly following~\cite{Ashwinkumar:2021kav}, and we fix the dictionary between Chern--Simons states and topology which is used throughout the paper.

\subsection{Theta functions, the Weil representation, and the Siegel--Weil formula}
Let $(\Lambda,Q)$ be an even integral lattice of rank $p+q$, with $p$ positive and $q$ negative eigenvalues and signature $\sigma:=p-q$. We write $Q(\ell):=Q(\ell,\ell)$, $\Lambda^*$ for the dual lattice and
\begin{equation}
  \Ds:=\Lambda^*/\Lambda
  \label{eq:disc-form}
\end{equation}
for the discriminant group~\cite{Wall1963,Nikulin1979}, which is equipped with the quadratic form $q$ and the bilinear form $(\cdot,\cdot)$ induced by $Q$:
\begin{equation}
  q(x):=\tfrac12Q(x,x)\bmod 1,\qquad (x,y):=Q(x,y)\bmod1\;.
  \label{eq:disc-form-pairing}
\end{equation}
The Gauss--Milgram formula~\cite{Milgram1974,MilnorHusemoller1973} reads
\begin{equation}
  \sum_{x\in\Ds}\e\bigl(q(x)\bigr)=\sqrt{|\Ds|}\;\e(\sigma/8)\;.
  \label{eq:milgram}
\end{equation}

\paragraph{Theta functions and the Weil representation.}
We identify $O(p,q)/(O(p)\times O(q))$ with the Grassmannian of $p$-dimensional subspaces $m\subset\Lambda\otimes\RR$ on which $Q$ is positive definite. The group $O(p,q)$ acts on it from the left by $g\cdot m=g(m)$. Writing $\ell=\ell_m+\ell_{m^\perp}$ for the orthogonal decomposition, $m$ splits $Q=Q_L-Q_R$ into the positive-definite pieces $Q_L(\ell):=Q(\ell_m)$ and $Q_R(\ell):=-Q(\ell_{m^\perp})$. The Siegel--Narain theta functions are
\begin{equation}
  \Theta_x(\tau;m):=\sum_{\ell\in\Lambda+x}\e\Bigl(\tfrac{\tau}{2}Q_L(\ell)-\tfrac{\bar\tau}{2}Q_R(\ell)\Bigr),\qquad x\in\Ds\;,
  \label{eq:theta-def}
\end{equation}
which we collect into $\Theta(\tau;m):=\sum_x\Theta_x(\tau;m)\,\ket{x}\in\CC[\Ds]$.

\paragraph{Moduli space with marked discriminant group.}
Let $O_Q(\ZZ)$ be the group of automorphisms of $\Lambda$ preserving $Q$, acting on the Grassmannian as above. An element $h\in O_Q(\ZZ)$ preserves $\Lambda^*$ and induces $\bar h\in\Aut(\Ds,q)$ by $\bar h(x+\Lambda)=hx+\Lambda$. Since $h$ is an isometry with $h(m)^\perp=h(m^\perp)$, we have $\ell_{h(m)}=h\bigl((h^{-1}\ell)_m\bigr)$, hence $Q_L^{h(m)}(\ell)=Q_L^{m}(h^{-1}\ell)$ and $Q_R^{h(m)}(\ell)=Q_R^{m}(h^{-1}\ell)$. Substituting $\ell'=h^{-1}\ell$, which runs over $h^{-1}(\Lambda+x)=\Lambda+\bar h^{-1}x$, gives
\begin{equation}
  \Theta_x(\tau;h\cdot m)=\Theta_{\bar h^{-1}x}(\tau;m)\;,\qquad\text{i.e.}\qquad \Theta(\tau;h\cdot m)=P(\bar h)\,\Theta(\tau;m)\;,
  \label{eq:theta-relabel}
\end{equation}
where $P(\bar h)\ket x:=\ket{\bar hx}$ is the permutation operator. Hence an individual component $\Theta_x$, and more generally a functional $\Psi(\Theta)$ with $\Psi\in\CC[\Ds]^*$, is not in general a function on the usual Narain moduli space
\begin{align}
  \cM:=O_Q(\ZZ)\backslash O(p,q)/(O(p)\times O(q))\;.
\end{align} 
We therefore work throughout on the finite cover
\begin{equation}
  \cM_Q:=O_Q(\ZZ)^{\rm disc}\backslash O(p,q)/(O(p)\times O(q)),\qquad
  O_Q(\ZZ)^{\rm disc}:=\ker\bigl(O_Q(\ZZ)\to\Aut(\Ds,q)\bigr)\;,
  \label{eq:MQ}
\end{equation}
on which the labels $x\in\Ds$ are fixed, and on which every $\Theta_x(\tau;\cdot)$ is a function. The subgroup $O_Q(\ZZ)^{\rm disc}$ has finite index, and we normalize the invariant measure on $\cM_Q$ to total volume one. For an observable invariant under a larger group, such as $\Theta_x$ under the stabilizer of $x$ used in~\cite{Ashwinkumar:2023jtz}, or $\Psi(\Theta)$ with $\Psi$ invariant under $\Aut(\Ds,q)$, the normalized average on $\cM_Q$ coincides with that on the corresponding smaller quotient, since the measure on a finite cover pushes forward to the normalized measure. For several Narain theories with independent moduli we use $(\cM_Q)^n$, i.e.\ all $n$ theories carry the \emph{same} marking of $\Ds$. This is needed because, for instance, $\sum_x\prod_i\Theta_x(\tau_i;m_i)$ is invariant under simultaneous relabelings $x\mapsto\bar hx$, but not under independent relabelings of the individual $m_i$.

\paragraph{Modular properties.}
With the automorphy factor
\begin{equation}
  j_{p,q}(\gamma,\tau):=(c\tau+d)^{p/2}(c\bar\tau+d)^{q/2},\qquad \gamma=\begin{pmatrix}a&b\\c&d\end{pmatrix}\;,
  \label{eq:j-def}
\end{equation}
the theta vector transforms as
\begin{equation}
  \Theta(\gamma\tau;m)=j_{p,q}(\gamma,\tau)\,\rho(\gamma)\,\Theta(\tau;m)\;,
  \label{eq:theta-mod}
\end{equation}
where $\rho$ is the Weil representation of $\SL(2,\ZZ)$. When $p+q$ is odd, $\rho$ is a representation of the double cover $\Mp(2,\ZZ)$ and $j_{p,q}$ requires a choice of branch, and we suppress the resulting signs throughout. Explicitly, on the generators $T:=\bigl(\begin{smallmatrix}1&1\\0&1\end{smallmatrix}\bigr)$ and $S:=\bigl(\begin{smallmatrix}0&-1\\1&0\end{smallmatrix}\bigr)$ of $\SL(2,\ZZ)$, which act by $T\tau=\tau+1$ and $S\tau=-1/\tau$,
\begin{equation}
  \rho(T)\ket{x}=\e\bigl(q(x)\bigr)\ket{x},\qquad
  \rho(S)\ket{x}=\frac{\e(-\sigma/8)}{\sqrt{|\Ds|}}\sum_{y\in\Ds}\e\bigl(-(x,y)\bigr)\ket{y}\;.
  \label{eq:weil}
\end{equation}
These formulas agree with~\cite[(20)]{Ashwinkumar:2021kav}. The physical partition function is
\begin{equation}
  Z_x:=\frac{\Theta_x}{\eta^p\bar\eta^q}\;.
  \label{eq:Zx}
\end{equation}
Since $\eta(\tau+1)=\e(1/24)\,\eta(\tau)$, the $T$-transformation of $Z_x$ contains, in addition to $\rho(T)$, the anomalous phase $\e(-\mathsf c/24)$, where $\mathsf c=\sigma$ is the chiral central charge. This is the framing anomaly of the Chern--Simons states. The phase is absent for $\Theta$, so we work with $\Theta$ throughout. The $\eta$-factors can be restored at the end as in~\cite[(45),(60)]{Ashwinkumar:2021kav}.

\paragraph{Degree-one Siegel--Weil and lens spaces.}
Since $\rho(T)\ket0=\ket0$, the Eisenstein vector
\begin{equation}
  E(\tau):=\sum_{\gamma\in\Ginf\backslash\SL(2,\ZZ)}j_{p,q}(\gamma,\tau)^{-1}\rho(\gamma)^{-1}\ket{0}
  \label{eq:E-def}
\end{equation}
is well defined, where $\Ginf:=\{\pm T^k\mid k\in\ZZ\}$. The summand $j_{p,q}(\gamma,\tau)^{-1}\rho(\gamma)^{-1}\ket0$ is invariant under $\gamma\mapsto T^k\gamma$, and for even $p+q$ also under $\gamma\mapsto-\gamma$, because $\rho(-\mathbbm 1)\ket0=\e(-\sigma/4)\ket0$ is compensated by $j_{p,q}(-\gamma,\tau)=\e(-\sigma/4)j_{p,q}(\gamma,\tau)$ (with principal branches, $c>0$). The factor $\rho(\gamma)^{-1}\ket0$ alone is not invariant under $\gamma\mapsto-\gamma$ when $\sigma\equiv2\bmod4$. The series converges absolutely for $p+q>4$ and transforms like $\Theta$. Indeed, the substitution $\gamma\mapsto\gamma\delta$ shows that $E(\delta\tau)=j_{p,q}(\delta,\tau)\rho(\delta)E(\tau)$. Moreover, for $a\in\Aut(\Ds,q)$ the permutation $P(a)\ket x:=\ket{ax}$ commutes with $\rho$ (by~\eqref{eq:weil}) and fixes $\ket0$, so that
\begin{equation}
  P(a)E(\tau)=E(\tau),\qquad\text{i.e.}\qquad E_{ax}(\tau)=E_x(\tau)\quad\text{for all }a\in\Aut(\Ds,q)\;;
  \label{eq:E-aut}
\end{equation}
in particular $E_{-x}=E_x$. We represent $\Ginf\backslash\SL(2,\ZZ)$ by coprime pairs $(c,d)$ with $c>0$, together with $(c,d)=(0,1)$, and $\gamma_{c,d}$ denotes a matrix with this lower row and with these signs. All Gauss sums $\varepsilon_{Q,x}$, and the amplitudes built from them below, are evaluated on these signed representatives. The components
\begin{equation}
  \varepsilon_{Q,x}(c,d):=\bra{x}\rho(\gamma_{c,d})^{-1}\ket{0},\qquad
  E_x(\tau)=\sum_{(c,d)}\frac{\varepsilon_{Q,x}(c,d)}{j_{p,q}(\gamma_{c,d},\tau)}
  \label{eq:eps-def}
\end{equation}
coincide with~\cite[(29)--(30)]{Ashwinkumar:2021kav}. Explicitly, $\varepsilon_{Q,x}(0,1)=\delta_{x,0}$ and, for $c>0$,
\begin{equation}
  \varepsilon_{Q,x}(c,d)=\frac{\e(\sigma/8)}{\sqrt{|\Ds|}\,c^{(p+q)/2}}\sum_{\ell\in\Lambda/c\Lambda}\e\Bigl(-\frac{d}{2c}Q(\ell+x)\Bigr)\;.
  \label{eq:eps}
\end{equation}

For $pq\neq0$ and $p+q>4$, the degree-one Siegel--Weil formula~\cite{Siegel1951,Weil1965} (see~\cite{Ashwinkumar:2021kav,Maloney:2020nni,Afkhami-Jeddi:2020ezh} for the physics application) states that
\begin{equation}
  \vev{\Theta(\tau;m)}_{\cM_Q}=E(\tau)\;,
  \label{eq:SW1}
\end{equation}
where $\vev{\cdot}_{\cM_Q}$ denotes the average with respect to the normalized invariant measure on the marked moduli space~\eqref{eq:MQ}. Component by component, this agrees with the averages of~\cite{Ashwinkumar:2021kav,Ashwinkumar:2023jtz} by the remark below~\eqref{eq:MQ}.

\subsection{Topological dictionary}
\label{sec:dict}
Let $V=D^2\times S^1$ be a solid torus with meridian $\mathfrak m$, and let $\mathfrak l$ be the longitude. We use the following dictionary between Chern--Simons states and topology.
\begin{enumerate}
  \item[(a)] $\ket{x}\in\cH(\partial V)=\CC[\Ds]$ is the state of $V$ with a Wilson line of charge $x$ along its core. Equivalently, $x$ is the holonomy around $\mathfrak m$.
  \item[(b)] $\rho(T)$ is the Dehn twist along $\mathfrak m$. By~(a) it is diagonal in this basis.
  \item[(c)] $\rho(\gamma)^{-1}\ket{0}$ is the complex conjugate of the state of a solid torus whose meridian is attached along the unoriented slope $c\mathfrak l+d\mathfrak m$. Equivalently, it is the state of this solid torus with the opposite orientation.
\end{enumerate}
Item (c) is consistent with $\Ginf$-invariance: the change of coset representative $\gamma\mapsto T^k\gamma$ preserves the lower row $(c,d)$, hence the slope, and $\rho(T^k\gamma)^{-1}\ket0=\rho(\gamma)^{-1}\rho(T)^{-k}\ket0=\rho(\gamma)^{-1}\ket0$. The right action $\gamma\mapsto\gamma T^k$, on the other hand, sends $(c,d)\mapsto(c,d+kc)$, i.e.\ it changes the slope, and $\rho(\gamma T^k)^{-1}\ket0=\rho(T)^{-k}\rho(\gamma)^{-1}\ket0$ is the effect of the twist $\rho(T)^{-k}$ along $\mathfrak m$ by item~(b). This is the origin of the Seifert moves in Section~\ref{sec:seifert}.

Item~(c) also reproduces the identification of $\varepsilon_{Q,x}(c,d)$ with the lens space $L(c,d)$ in~\cite{Ashwinkumar:2021kav}. Indeed, gluing the solid torus of item~(a) to one whose meridian is $c\mathfrak l+d\mathfrak m$ gives $L(c,d)$, with $|H_1|=|\mathfrak m\cdot(c\mathfrak l+d\mathfrak m)|=c$. The complex conjugation in (c) is the one noted below~\cite[(58)]{Ashwinkumar:2021kav}, and it is confirmed by the comparison with the surgery formula in Section~\ref{sec:ZMcomparison}.

To avoid writing complex conjugates throughout, \emph{we denote by $Z_{\rm CS}[M]$ the partition function of $M$ with the orientation reversed}, i.e.\ $Z_{\rm CS}[M]:=Z^{\rm std}_{\rm CS}[-M]=\overline{Z^{\rm std}_{\rm CS}[M]}$, where $Z^{\rm std}_{\rm CS}$ is the partition function for the given orientation. Throughout, $\doteqf$ denotes equality up to an overall framing phase (a power of $\e(\sigma/8)$).

In this language, \eqref{eq:intro-adkly} can be read as follows. The vector $\Theta(\tau;m)$ is the Narain theory placed on $\partial V$. Averaging replaces it by a sum over solid tori $V_{(c,d)}$ glued in with all possible slopes. Pairing with $\bra{x}$, the solid torus with meridian $\mathfrak m$ and Wilson line $x$, closes the manifold into a lens space.

\section{Seifert manifolds and their Abelian Chern--Simons invariants}
\label{sec:seifert}
In this section we compute the Abelian Chern--Simons partition functions of Seifert manifolds by cutting them along the interface tori of~\eqref{eq:intro-decomp}. The result, equation~\eqref{eq:ZM} below, is the input for the ensemble averages of the following sections.

\subsection{Seifert data and marked presentations}
Fix a closed oriented surface $\Sigma_g$ and integers $n\geq0$ and $\chi\in\ZZ$. Let $\Sigma_{g,n+1}$ be $\Sigma_g$ with $n+1$ discs removed, and let
\begin{equation}
  \Mcore_{g,n+1}:=\Sigma_{g,n+1}\times S^1\;.
\end{equation}
On each boundary torus $\mathbb T_i$ ($i=0,\dots,n$) we choose the basis $(\mathfrak l_i,\mathfrak m_i)$, where $\mathfrak l_i$ is the boundary of the section $\Sigma_{g,n+1}\times\{\textrm{pt}\}$ and $\mathfrak m_i$ is the regular fiber. Given coprime pairs $(\alpha_i,\beta_i)$ with $\alpha_i\ge0$, we define~\cite{Seifert1933,Orlik1972,Scott1983}
\begin{equation}
  M\bigl(g;\chi;(\alpha_1,\beta_1),\dots,(\alpha_n,\beta_n)\bigr)
  :=\Mcore_{g,n+1}\ \cup_{\mathbb T_0}V_0\ \cup_{\mathbb T_1}V_1\cup\dots\cup_{\mathbb T_n}V_n\;,
  \label{eq:seifert-def}
\end{equation}
where the meridian of $V_0$ is attached along $\mathfrak l_0+\chi \mathfrak m_0$ and the meridian of $V_i$ along $\alpha_i\mathfrak l_i+\beta_i\mathfrak m_i$. Note that the fiber $\mathfrak m_i$ is the meridian of $V_i$ only for the slope $(0,1)$. We orient $M$ by the orientations of the base and the fiber. When all $\alpha_i\ge1$, the (rational) Euler number of $M$ is
\begin{equation}
  \mathsf e:=\chi+\sum_{i=1}^n\frac{\beta_i}{\alpha_i}\;,
  \label{eq:euler}
\end{equation}
and $\chi$ is its integer contribution from $V_0$. With these conventions, $M(g;\chi;(\alpha_i,\beta_i))$ is the Seifert manifold $\{\chi;(o_1,g);(\alpha_1,\beta_1),\dots,(\alpha_n,\beta_n)\}$ of~\cite{Orlik1972}, whose Euler number in the convention of~\cite{Orlik1972} is $-\mathsf e$. Equivalently, $M$ is obtained by surgery on an unknot $K_0$ with coefficient $\chi$ and on $n$ meridians of $K_0$ with coefficients $-\alpha_i/\beta_i$, together with $g$ Borromean pairs of $0$-framed components (see Figure~\ref{fig:surgery}, and Appendix~\ref{sec:ZMcanderiv} for the definition of surgery).

The nature of the $i$-th filling depends on $\alpha_i$. For $\alpha_i\ge2$ it gives an exceptional fiber of type $(\alpha_i,\beta_i)$, and for $\alpha_i=1$ it gives a regular fiber, which can be absorbed into $\chi$. The case $\alpha_i=0$ is called a \emph{degenerate filling}: the fiber bounds a disc, and the result is independent of $\chi$. Indeed, filling one boundary torus of $\Sigma_{g,n+1}\times S^1$ along the fiber gives $\#^{2g}(S^2\times S^1)$ with $n$ solid tori removed, whose core circles lie in disjoint balls, and each fiber $\mathfrak m_i$ bounds a disc in the complement. Filling the $i$-th boundary torus along $\alpha_i\mathfrak l_i+\beta_i\mathfrak m_i$ therefore produces a connected summand $L(\alpha_i,\beta_i)$, with $|H_1|=|\mathfrak m_i\cdot(\alpha_i\mathfrak l_i+\beta_i\mathfrak m_i)|=\alpha_i$. This is $S^2\times S^1$ for a further degenerate filling and $S^3$ for $\alpha_i=1$. Hence, if $k\ge1$ of the fillings are degenerate,
\begin{equation}
  M\cong\#^{\,2g+k-1}(S^2\times S^1)\ \#\ \mathop{\#}_{\alpha_i\ge1}L(\alpha_i,\beta_i)\;,
  \label{eq:degenerate}
\end{equation}
where summands $L(1,\beta)\cong S^3$ are trivial. This is in general a connected sum and need not be a Seifert manifold. Among reducible orientable 3-manifolds, the only Seifert manifolds are $S^2\times S^1$ and $\mathbb{RP}^3\#\mathbb{RP}^3$, the latter fibered over a non-orientable base~\cite[Prop.~1.13 and Section~2.1]{Hatcher3M}. The second case does occur here, for $g=0$, one degenerate filling and the two fillings $(2,1),(2,1)$. We keep the degenerate fillings because they are the analogue of the identity-coset term $\delta_{x,0}$ of $E_x$, i.e.\ the term with $(c,d)=(0,1)$, and modular invariance requires them (Section~\ref{sec:egf}).

Using the dictionary of Section~\ref{sec:dict} with $(\mathfrak l,\mathfrak m)=(\mathfrak l_i,\mathfrak m_i)$ on $\mathbb T_i$, the slope $(\alpha_i,\beta_i)$ corresponds to $(c_i,d_i)$. A \emph{labeled marked presentation} is therefore
\begin{equation}
  \tilde{\mathfrak s}:=\bigl(g;\chi;\gamma_1,\dots,\gamma_n\bigr),\qquad \gamma_i\in\Ginf\backslash\SL(2,\ZZ)\;.
\end{equation}
Two kinds of redundancy relate presentations of the same manifold. The first are the \emph{Seifert moves} ($\ZZ^n$), which in our conventions read
\begin{equation}
  M\bigl(g;\chi;\dots,(\alpha_i,\beta_i+k\alpha_i),\dots\bigr)\cong M\bigl(g;\chi+k;\dots,(\alpha_i,\beta_i),\dots\bigr)\;,
  \label{eq:seifert-move}
\end{equation}
that is, $(\chi;\gamma_i)\sim(\chi-k;\gamma_iT^{k})$. The second is \emph{relabeling} ($\mathfrak S_n$), i.e.\ permuting the exceptional fibers. For $\alpha_i\ge2$, each $\ZZ^n$-orbit contains a unique \emph{normalized} representative with $0<\beta_i<\alpha_i$.

\subsection{The core functional}
\label{sec:core}
The Chern--Simons theory with kinetic matrix $Q$~\cite{Belov:2005ze,Lu:2012dt,Ashwinkumar:2021kav} assigns $\CC[\Ds]$ to each torus.\footnote{For Chern--Simons theory on Seifert manifolds in general, see e.g.~\cite{Freed:1991wd,Rozansky:1993zx,LawrenceRozansky1999,Blau:2013oha}.} The core with its $0$-th boundary filled by $V_0$ defines a functional
\begin{equation}
  Z_{\rm CS}\bigl[\Mcore_{g,n+1}\cup_{\mathbb T_0}V_0\bigr]\ \in\ \bigl(\CC[\Ds]^{\otimes n}\bigr)^*\;,
\end{equation}
which, like any Chern--Simons amplitude, is defined only up to a framing phase. We call the basis $\{\ket{x}_i\}$ of $\cH(\mathbb T_i)$ defined by item~(a) of Section~\ref{sec:dict} with $\mathfrak m=\mathfrak m_i$ the \emph{fiber basis}. In this basis $x_i$ is the holonomy around the regular fiber. Our first result is that this functional is diagonal in the fiber basis: $Z_{\rm CS}[\Mcore_{g,n+1}\cup_{\mathbb T_0}V_0]\doteqf\cV_{g,n;\chi}$, where we \emph{define}
\begin{equation}
  \cV_{g,n;\chi}:=|\Ds|^{\,g-1+\frac n2}\sum_{x\in\Ds}\e\bigl(-\chi\,q(x)\bigr)\,\bra{x}_1\otimes\cdots\otimes\bra{x}_n \;.
  \label{eq:core}
\end{equation}
In components,
\[
  \cV_{g,n;\chi}(x_1,\dots,x_n)=|\Ds|^{g-1+n/2}\,\e(-\chi q(x_1))\,\delta_{x_1=\cdots=x_n}\qquad(n\geq1)\;.
\]
From now on $\cV_{g,n;\chi}$ always denotes the right-hand side of~\eqref{eq:core}, and all equalities involving it are exact.

The form of~\eqref{eq:core} is easy to understand: all regular fibers are homologous in the core, so the fiber holonomy $x$ is the same on every boundary. The filling $V_0$ acts as $\chi$ Dehn twists along the fiber, which are diagonal by item~(b) of Section~\ref{sec:dict}. The normalization follows from the Verlinde formula, as we show at the end of this subsection.

\paragraph{Checks.}
For arbitrary $\Ds$, \eqref{eq:core} gives the expected results in simple cases. For $(g,n,\chi)=(0,0,0)$, i.e.\ $S^2\times S^1$, it gives $|\Ds|^{-1}\sum_x1=1$, and for $(0,0,1)$, i.e.\ $S^3$, it gives $|\Ds|^{-1}\sum_x\e(-q(x))\doteqf|\Ds|^{-1/2}=S_{00}$ by~\eqref{eq:milgram}. For $(g,0,0)$, i.e.\ $\Sigma_g\times S^1$, we obtain $|\Ds|^g=\dim\cH(\Sigma_g)$, and for $(0,0,\chi)$, i.e.\ $L(\chi,1)$, we obtain $|\Ds|^{-1}\sum_x\e(-\chi q(x))$, the expected quadratic Gauss sum. Moreover, gluing $\cV_{g_1,n_1;\chi_1}$ and $\cV_{g_2,n_2;\chi_2}$ along one leg, or self-gluing two legs, reproduces~\eqref{eq:core} with $(g,\chi)=(g_1+g_2,\chi_1+\chi_2)$ or $g\to g+1$, respectively, and the correct power of $|\Ds|$: the exponent satisfies $g-1+n/2=(g_1-1+n_1/2)+(g_2-1+n_2/2)$ with $n=n_1+n_2-2$.

\paragraph{Derivation.}\label{sec:corederiv} We now derive~\eqref{eq:core} in three steps.

\emph{Step 1: section basis.} Let $\ket{h}^{\rm sec}_i$ be the state of a solid torus whose meridian is $\mathfrak l_i$, with a Wilson line $h$ along its core (parallel to $\mathfrak m_i$). Filling all $n+1$ boundaries of $\Mcore_{g,n+1}$ in this way gives $\Sigma_g\times S^1$ with $n+1$ Wilson lines along fibers. The Verlinde formula~\cite{Verlinde:1988sn} for the Abelian theory, whose modular $S$-matrix is $S_{xy}=|\Ds|^{-1/2}\e(-(x,y))$, i.e.\ $\rho(S)$ of~\eqref{eq:weil} without the phase $\e(-\sigma/8)$, gives
\begin{equation}
  \begin{aligned}
    Z_{\rm CS}\bigl[\Sigma_g\times S^1;h_0,\dots,h_n\bigr] & =\sum_{y\in\Ds}S_{0y}^{\,2-2g-(n+1)}\prod_{i=0}^nS_{h_iy}                                                                  \\
                                                  & =|\Ds|^{g-1}\sum_{y\in\Ds}\e\Bigl(-\Bigl(\sum_ih_i,\,y\Bigr)\Bigr)=|\Ds|^{g}\,\delta\Bigl(\sum_ih_i=0\Bigr)\;.
  \end{aligned}
\end{equation}

\emph{Step 2: change to the fiber basis.} The section-basis state $\ket{h}^{\rm sec}_i$ is the state of a solid torus whose meridian is $\mathfrak l_i$, i.e.\ the slope $(c,d)=(1,0)$ in the dictionary of Section~\ref{sec:dict} with $(\mathfrak l,\mathfrak m)=(\mathfrak l_i,\mathfrak m_i)$, carrying a Wilson line $h$. By item~(c) (with $\gamma=S$) and Wilson-line transport, $\langle x|h\rangle^{\rm sec}=\bigl(\rho(S)^{-1}\bigr)_{xh}=\e(\sigma/8)|\Ds|^{-1/2}\e\bigl((x,h)\bigr)$. The overall phases are powers of $\e(\sigma/8)$, absorbed in $\doteqf$. The sign of $(x,h)$ is also immaterial, since $\sum_{h:\,\sum h_i=0}\prod_i\e(\pm(x_i,h_i))$ is the same for both signs. Fourier transforming each of the $n+1$ legs gives
\begin{equation}
  |\Ds|^{g}\,|\Ds|^{-(n+1)/2}\sum_{h:\ \sum h_i=0}\prod_i\e\bigl((x_i,h_i)\bigr)=|\Ds|^{g-1+(n+1)/2}\,\delta_{x_0=x_1=\cdots=x_n}\;.
\end{equation}

\emph{Step 3: the filling $V_0$.} Fill $\mathbb T_0$ with slope $(1,\chi)$, i.e.\ with $\rho(\gamma_{1,\chi})^{-1}\ket0$ and $\gamma_{1,\chi}=ST^\chi$. Then $\langle x|\rho(\gamma_{1,\chi})^{-1}|0\rangle=\e(-\chi q(x))\,\e(\sigma/8)|\Ds|^{-1/2}$. Contracting the $x_0$ leg removes one factor of $|\Ds|^{1/2}$, which gives~\eqref{eq:core}.

Since gluing two cores along one leg via the cylinder $\sum_x\ket x\otimes\ket x$ reproduces~\eqref{eq:core}, as noted above, the result does not depend on how $\Sigma_g$ is decomposed into pairs of pants.

\subsection{Abelian Chern--Simons partition functions of Seifert manifolds}
Filling the $n$ remaining boundaries with the states $\rho(\gamma_i)^{-1}\ket0$ and using~\eqref{eq:core} gives the Chern--Simons partition function of $M(g;\chi;(\alpha_i,\beta_i))$ up to a framing phase. Since this phase depends on the presentation and not only on the manifold (see below), we define, for a labeled marked presentation $\tilde{\mathfrak s}=(g;\chi;\gamma_1,\dots,\gamma_n)$, the \emph{presentation-framed amplitude}
\begin{equation}
  Z^{\rm pr}_{\rm CS}\bigl[g;\chi;\gamma_1,\dots,\gamma_n\bigr]:=\cV_{g,n;\chi}\Bigl(\bigotimes_{i=1}^n\rho(\gamma_i)^{-1}\ket0\Bigr)=|\Ds|^{\,g-1+\frac n2}\sum_{x\in\Ds}\e\bigl(-\chi q(x)\bigr)\prod_{i=1}^n\varepsilon_{Q,x}(\alpha_i,\beta_i)
  \label{eq:ZM}
\end{equation}
where $(\alpha_i,\beta_i)$ is the lower row of $\gamma_i$. This is an exact definition, and
\begin{equation}
  Z^{\rm pr}_{\rm CS}\bigl[g;\chi;\gamma_1,\dots,\gamma_n\bigr]\doteqf Z_{\rm CS}\bigl[M(g;\chi;(\alpha_i,\beta_i)_{i=1}^n)\bigr]\;.
  \label{eq:ZMfr}
\end{equation}
Note that the right-hand side of~\eqref{eq:ZM} is exactly invariant under the Seifert moves~\eqref{eq:seifert-move}, since
\begin{equation*}
  \varepsilon_{Q,x}(\alpha,\beta+k\alpha)=\e\bigl(-kq(x)\bigr)\,\varepsilon_{Q,x}(\alpha,\beta)\;,
\end{equation*}
and under $\mathfrak S_n$, since~\eqref{eq:core} is symmetric.

A regular fiber $(1,d)$ contributes $\varepsilon_{Q,x}(1,d)=\e(\sigma/8)|\Ds|^{-1/2}\e(-dq(x))$, so that filling one leg of $\cV_{g,n+1;\chi}$ with $(1,d)$ gives $\e(\sigma/8)\,\cV_{g,n;\chi+d}$. Note that the two presentations describe the same manifold, but their amplitudes~\eqref{eq:ZM} differ by the phase $\e(\sigma/8)$. Hence the phase in~\eqref{eq:ZMfr} cannot be dropped in a sum over presentations. For instance, for $\Ds=\ZZ_3$ and $\sigma=2$ (Section~\ref{sec:example}), the two presentations $M(0;1;\emptyset)$ and $M(0;0;(1,1))$ of $S^3$ give $Z^{\rm pr}_{\rm CS}=-i/\sqrt3$ and $1/\sqrt3$, respectively. The precise relation to the canonically framed invariant is~\eqref{eq:ZMcan} below. A degenerate filling $(0,1)$ contributes $\delta_{x,0}$ and removes the $\chi$-dependence, so that~\eqref{eq:ZM} gives $|\Ds|^{g-1+n/2}\prod_{\alpha_i\ge1}\varepsilon_{Q,0}(\alpha_i,\beta_i)$. Since $\varepsilon_{Q,0}(\alpha,\beta)\doteqf Z_{\rm CS}[L(\alpha,\beta)]$ and $Z_{\rm CS}[M_1\#M_2]=Z_{\rm CS}[M_1]Z_{\rm CS}[M_2]/Z_{\rm CS}[S^3]$, this is the partition function of the connected sum~\eqref{eq:degenerate}.

\paragraph{Comparison with surgery formulas.}\label{sec:ZMcomparison}
Equation~\eqref{eq:ZM} can be compared with the standard formula for Abelian Chern--Simons invariants~\cite{MOO1992}, \cite[(3) and Appendix~A]{Deloup1999}. Let $M$ be obtained by integral surgery (recalled in Appendix~\ref{sec:ZMcanderiv}) on a framed link with $N$ components. We denote by $W$ its linking matrix, whose diagonal entries $W_{ii}$ are the framings of the components and whose off-diagonal entries $W_{ij}$ are their linking numbers, and by $\sigma(W)$ the signature of $W$ (for Seifert manifolds, see the explicit form of $W$ in Appendix~\ref{sec:ZMcanderiv}). For the given orientation of $M$,
\begin{equation}
  Z^{\rm can}_{\rm CS}[M]=|\Ds|^{-\frac{N+1}{2}}\,\e\Bigl(-\frac{\sigma\,\sigma(W)}{8}\Bigr)\sum_{x\in\Ds^N}\e\Bigl(\sum_{i}W_{ii}\,q(x_i)+\sum_{i<j}W_{ij}\,(x_i,x_j)\Bigr)\;,
  \label{eq:surgery}
\end{equation}
normalized so that $Z^{\rm can}_{\rm CS}[S^3]=|\Ds|^{-1/2}$ and $Z^{\rm can}_{\rm CS}[S^2\times S^1]=1$. We call the framing with this choice of phase, fixed by the signature term, the \emph{canonical framing}. The invariant~\eqref{eq:surgery} depends only on $b_1(M)$ and the linking form~\cite[Theorem~1]{Deloup1999}. For $b_1(M)=0$, reciprocity for Gauss sums~\cite[Theorem~3]{Deloup1999}, \cite{DeloupTuraev2007} turns it into a Gauss sum of the linking form on $H:=H_1(M)$, tensored with the lattice. Here $H$ is generated by the meridians $\mathfrak m_{K_i}$ of the link components, and the linking form is $\lambda_M(\mathfrak m_{K_i},\mathfrak m_{K_j})=-(W^{-1})_{ij}\bmod1$. For instance, for $p$-surgery on the unknot, $p\mathfrak m$ bounds the meridian disc of the glued solid torus minus a Seifert disc of the unknot, which gives $\lambda_M(\mathfrak m,\mathfrak m)=-1/p$. Then
\begin{equation}
  Z^{\rm can}_{\rm CS}[M]=|\Ds|^{-\frac12}\,|H|^{-\frac{p+q}{2}}\sum_{y\in H\otimes\Lambda}\e\Bigl(\tfrac12(\lambda_M\otimes Q)(y,y)\Bigr)\;.
  \label{eq:linking}
\end{equation}
As noted in Section~\ref{sec:seifert}, $M(g;\chi;(\alpha_i,\beta_i))$ is obtained by surgery on an unknot $K_0$ with coefficient $\chi$, on $n$ meridians of $K_0$ with rational coefficients $-\alpha_i/\beta_i$ (each expanded into a chain of integral surgeries by a continued fraction), and on $g$ Borromean pairs of $0$-framed components, which do not contribute to the linking matrix. The orientation-reversed manifold $-M$ is the mirror surgery $(-\chi;\alpha_1/\beta_1,\dots,\alpha_n/\beta_n)$. Evaluating~\eqref{eq:surgery} for this presentation gives, for even $p+q$,
\begin{equation}
  \overline{Z^{\rm can}_{\rm CS}\bigl[M(g;\chi;(\alpha_i,\beta_i)_{i=1}^n)\bigr]}=\e\Bigl(-\frac{\sigma}{8}\bigl(n-\operatorname{sign}(\mathsf e)\bigr)\Bigr)\,Z^{\rm pr}_{\rm CS}\bigl[g;\chi;\gamma_1,\dots,\gamma_n\bigr]\;,
  \label{eq:ZMcan}
\end{equation}
where $\mathsf e$ is the Euler number~\eqref{eq:euler}, $\operatorname{sign}(0)=0$, and all $\alpha_i\ge1$. The left-hand side depends only on $M$, while $n$ and $\mathsf e$ are invariant under Seifert moves and relabeling. Adding a regular fiber raises $n$ by one, which compensates the phase $\e(\sigma/8)$ found above; in the $\ZZ_3$ example, \eqref{eq:ZMcan} gives $1/\sqrt3$ for both presentations of $S^3$. Equation~\eqref{eq:ZMcan} is derived in Appendix~\ref{sec:ZMcanderiv} by evaluating~\eqref{eq:surgery} leg by leg: the Gauss sum over the chain attached to the $i$-th meridian reproduces $\bra{x}\rho(\gamma_i)^{-1}\ket0$, and the phase is fixed by Sylvester's law of inertia together with a parity property of continued fractions.

Equation~\eqref{eq:ZMcan} fixes the framing phase left unspecified in~\eqref{eq:ZMfr}. It also fixes the orientation conventions: the amplitude~\eqref{eq:ZM} is, up to this phase, the partition function of $-M$, i.e.\ the complex conjugate of that of $M$, in accordance with item~(c) of Section~\ref{sec:dict} and with~\cite[(58)]{Ashwinkumar:2021kav}. Moreover, for even $p+q$ the representation $\rho$ is a genuine representation of $\SL(2,\ZZ)$, so no Rademacher-type phases appear, in contrast to the non-Abelian lens-space invariants of~\cite{Jeffrey:1992tk}. The entire framing dependence is the power of $\e(\sigma/8)$ in~\eqref{eq:ZMcan}. For lens spaces ($g=0$, $n\le2$), \eqref{eq:ZMcan} gives the canonically framed version of the identification of $\varepsilon_{Q,x}$ with lens-space partition functions in~\cite{Ashwinkumar:2021kav}.

To summarize, $Z_{\rm CS}[M]$ is the partition function of $-M$, defined up to a framing phase; $Z^{\rm pr}_{\rm CS}$ is its exact representative~\eqref{eq:ZM} in the framing fixed by the presentation; and $Z^{\rm can}_{\rm CS}[M]$ is the partition function of $M$ itself in the canonical framing, which is why its complex conjugate appears in~\eqref{eq:ZMcan}.

\subsection{Non-orientable bases}
\label{sec:nonorientable}
So far the base $\Sigma_g$ has been orientable. Seifert manifolds with orientable total space can also have a non-orientable base $N_h$, the connected sum of $h\ge1$ projective planes; the fiber is then reversed along every orientation-reversing loop of the base. Such bases arise from the self-gluings of Section~\ref{sec:selfglue}. Since the fiber is reversed, the fiber holonomy $x$ is identified with $-x$, and the core functional is supported on the $2$-torsion subgroup $\Ds[2]:=\{x\in\Ds:2x=0\}$. The crosscap is the twisted $I$-bundle over the Klein bottle, $K\tilde\times I$, which is Seifert fibered over the M\"obius band.\footnote{Twisted $I$-bundles have recently appeared among the geometries generated by surgery moves in pure three-dimensional gravity~\cite{Belin:2026pko}. See also~\cite{Yin:2007at,Maxfield:2016mwh}.} In the fiber basis its Chern--Simons functional is
\begin{equation}
  \Psi_{K\tilde\times I}\doteqf\sum_{x\in\Ds[2]}\nu_x\,\bra{x},\qquad \nu_x:=\e\bigl(2q(x)\bigr)=\pm1\;,
  \label{eq:crosscap}
\end{equation}
where $\nu_x$ is the Frobenius--Schur indicator of the self-conjugate anyon $x$, the analogue~\cite{Bantay:1996zk} of the classical indicator for finite groups~\cite{FrobeniusSchur1906} (Appendix~\ref{app:nonor}). Gluing $h$ copies of~\eqref{eq:crosscap} into the core gives the core functional for the base $N_{h,n+1}$, the surface $N_h$ with $n+1$ discs removed:
\begin{equation}
  \cV^{N}_{h,n;\chi}:=|\Ds|^{\,\frac h2-1+\frac n2}\sum_{x\in\Ds[2]}\nu_x^{\,h}\,\e\bigl(-\chi q(x)\bigr)\,\bra{x}_1\otimes\cdots\otimes\bra{x}_n
  \label{eq:core-nonor}
\end{equation}
which, as for~\eqref{eq:core}, we take as the exact definition of $\cV^{N}_{h,n;\chi}$, and it equals the Chern--Simons functional up to a framing phase. The power of $|\Ds|$ is again $-\tfrac12$ times the Euler characteristic $2-h-n$ of the punctured base. The results of Section~\ref{sec:main} extend to non-orientable bases by replacing $\cV_{g,n;\chi}$ with $\cV^{N}_{h,n;\chi}$. In particular, the Seifert--Eisenstein series becomes
\begin{equation}
  \cZ^{N}_{h,n;\chi}(\btau)=|\Ds|^{\,\frac h2-1+\frac n2}\sum_{x\in\Ds[2]}\nu_x^{\,h}\,\e\bigl(-\chi q(x)\bigr)\prod_{i=1}^nE_x(\tau_i)\;,
  \label{eq:Z-nonor}
\end{equation}
in which only the $2$-torsion components $E_x$ enter. For $|\Ds|$ odd only $E_0$ appears.

\section{The Seifert--Eisenstein series}
\label{sec:main}
We now place the Narain theories on the interface tori and average over their moduli. In this section the moduli are taken to be independent, so that only the degree-one Siegel--Weil formula is needed; the case of a common modulus is treated in Section~\ref{sec:common}. We state the results for even $p+q$, where $\rho$ is a genuine representation of $\SL(2,\ZZ)$ and no branch choices are needed. For odd $p+q$, \eqref{eq:main} below continues to hold with the same branch conventions on both sides, but the comparison~\eqref{eq:ZMcan} with the canonically framed invariants requires the metaplectic phases (Section~\ref{sec:discussion}).

\subsection{The replica average}
\label{sec:replica}
For $\btau:=(\tau_1,\dots,\tau_n)\in\HH^n$ we define the (labeled) \emph{Seifert--Eisenstein series}
\begin{equation}
  \cZ_{g,n;\chi}(\btau):=\sum_{\gamma_1,\dots,\gamma_n\in\Ginf\backslash\SL(2,\ZZ)}
  Z^{\rm pr}_{\rm CS}\bigl[g;\chi;\gamma_1,\dots,\gamma_n\bigr]\ \prod_{i=1}^n j_{p,q}(\gamma_i,\tau_i)^{-1}\;,
  \label{eq:Zdef}
\end{equation}
where $Z^{\rm pr}_{\rm CS}$ is the presentation-framed amplitude~\eqref{eq:ZM} with $(\alpha_i,\beta_i)=(c_i,d_i)$, i.e.\ the Chern--Simons partition function of $M(g;\chi;\gamma_1,\dots,\gamma_n)$ (with reversed orientation) in the framing fixed by the presentation. Its relation to the canonical framing is~\eqref{eq:ZMcan}. The sum runs over labeled \emph{marked} presentations, including regular and degenerate fillings, and not over homeomorphism classes. Each Seifert manifold therefore appears many times, related by Seifert moves and relabeling, and possibly also by inequivalent Seifert data describing homeomorphic manifolds (see Section~\ref{sec:normalized}). The terms with at least one degenerate filling are the connected sums~\eqref{eq:degenerate}, which in general are not Seifert manifolds. They come from the identity-coset terms $\delta_{x,0}$ of the $E_x$ and cannot be dropped without spoiling the modular properties (Section~\ref{sec:egf}). We will nevertheless continue to refer to~\eqref{eq:Zdef} as the Seifert sum.

The Narain observable that produces~\eqref{eq:Zdef} is
\begin{equation}
  \begin{aligned}
    \cO_{g,n;\chi}(\btau;m_1,\dots,m_n) & :=\cV_{g,n;\chi}\Bigl(\Theta(\tau_1;m_1)\otimes\cdots\otimes\Theta(\tau_n;m_n)\Bigr)  \\
                                        & =|\Ds|^{g-1+\frac n2}\sum_{x\in\Ds}\e(-\chi q(x))\prod_{i=1}^n\Theta_x(\tau_i;m_i)\;.
  \end{aligned}
  \label{eq:Odef}
\end{equation}
It describes $n$ replicas of the Narain theory, coupled only through the common fiber holonomy $x$ and weighted by the topological spin $\e(-\chi q(x))$. Our main result for independent moduli is the following.

\begin{proposition}\label{prop:main}
  Let $(\Lambda,Q)$ be an even lattice of signature $(p,q)$ with $p+q$ even, $pq\ne0$ and $p+q>4$. Let $\cM_Q$ be the moduli space with marked discriminant group~\eqref{eq:MQ}, and average the moduli $(m_1,\dots,m_n)\in(\cM_Q)^n$ independently with respect to the product of the normalized invariant measures. Then, for all $g\ge0$, $n\ge0$, $\chi\in\ZZ$ and $\btau\in\HH^n$,
  \begin{equation}
    \vev{\cO_{g,n;\chi}(\btau;m_1,\dots,m_n)}_{(\cM_Q)^n}
      =|\Ds|^{g-1+\frac n2}\sum_{x\in\Ds}\e(-\chi q(x))\prod_{i=1}^nE_x(\tau_i)
      =\cZ_{g,n;\chi}(\btau)\;,
    \label{eq:main}
  \end{equation}
  where $\cZ_{g,n;\chi}$ is the sum~\eqref{eq:Zdef} over labeled marked presentations, weighted by the presentation-framed amplitudes~\eqref{eq:ZM}. The sum~\eqref{eq:Zdef} converges absolutely.
\end{proposition}

\begin{proof}
  The first equality holds because the sum over $x$ is finite with $m$-independent coefficients, so the average over independent moduli factorizes and~\eqref{eq:SW1} applies to each factor. For the second, we insert~\eqref{eq:E-def} for each $E_x$. Since each $E_x$ converges absolutely, the finite sum over $x$ can be exchanged with the $n$-fold sum over $(\gamma_i)$, giving
  \[
    \sum_{\gamma_1,\dots,\gamma_n}\Bigl[|\Ds|^{g-1+\frac n2}\sum_x\e(-\chi q(x))\prod_i\varepsilon_{Q,x}(c_i,d_i)\Bigr]\prod_ij_{p,q}(\gamma_i,\tau_i)^{-1}\;.
  \]
  By~\eqref{eq:ZM} the bracket is exactly $Z^{\rm pr}_{\rm CS}[g;\chi;\bgamma]$. Absolute convergence of~\eqref{eq:Zdef} follows from that of each $E_x$, since the sum is a finite linear combination of products of absolutely convergent series.
\end{proof}

\paragraph{Interpretation.} In the language of symmetry-TFT (SymTFT) sandwiches~\cite{Freed:2022qnc}, each $\Theta(\tau_i;m_i)$ is the relative Narain theory of $\Lambda$ on the boundary of the bulk TQFT, and $\Theta_x$ is its sector in which the anyon $x$ threads the torus. At a generic point of $\cM_Q$ this boundary theory is irrational, so it is not obtained from a local boundary condition of the pure Chern--Simons theory; it can instead be realized in the topological limit of Maxwell--Chern--Simons theory, whose kinetic term carries the moduli~\cite{Gukov:2004id,Ashwinkumar:2021kav}. The core is a bulk TQFT amplitude joining the $n$ sandwiches: a single anyon $x$ runs along the fiber through all of them, and its world-line acquires the spin $\e(-\chi q(x))$ from the twisted filling $V_0$ and the factor $S_{0x}^{2-2g-n}=|\Ds|^{g-1+n/2}$ from the base. For $(g,n,\chi)=(0,2,0)$, $\cO$ is the diagonal pairing $\sum_x\Theta_x(\tau_1;m_1)\Theta_x(\tau_2;m_2)$ of two relative theories. Entangled boundary states of SymTFTs, and their relation to ensemble averages and $\alpha$-states, have been studied in~\cite{Torres:2025jcb}. For $g\ge1$ the extra factors $|\Ds|^{g}$ count the anyon lines that can run around the handles of the base. An interpretation of both the Narain and the Marolf--Maxfield averages as averages over topological boundary conditions of a SymTFT was developed in~\cite{Yu:2026gdf}.

\paragraph{Relation to the lens-space sum.}
Take $(g,n,\chi)=(0,2,0)$. Then $\cV_{0,2;0}=\sum_x\bra x\otimes\bra x$ is the identity cylinder $T^2\times I$. Freezing the second leg to the degenerate filling $\ket{h}$ (a solid torus with meridian $\mathfrak m$ and Wilson line $h$) gives back $E_h(\tau)$. The lens-space sum of~\cite{Ashwinkumar:2021kav} is therefore the Seifert sum over $S^2$ with one varying and one frozen fiber. Similarly, $(g,n)=(0,1)$ gives $|\Ds|^{-1/2}\sum_x\e(-\chi q(x))E_x(\tau)$, again a lens-space sum but with a different frozen fiber $(1,\chi)$. The Dedekind $\eta$-factors of the physical partition functions can be restored at the end; this is done in Appendix~\ref{app:eta}, where the Chern--Simons weights appear in the framing defined by the summed solid tori.

\paragraph{Positive-definite case.}
For $q=0$ the moduli space is a point. For even $p>4$, the Siegel--Weil formula then holds for the mass-weighted average over the genus of $Q$~\cite{Siegel1935,Ashwinkumar:2021kav}. The analogue of the marking~\eqref{eq:MQ} is as follows. The average runs over pairs $(L,\iota)$ of a lattice $L$ in the genus and an isometry $\iota:\Ds_L\to\Ds$, up to isomorphism. Each pair is weighted by $1/|O(L)^{\rm disc}|$, where $O(L)^{\rm disc}$ is the kernel of $O(L)\to\Aut(\Ds_L,q)$, and $\Theta_L$ is transported to $\CC[\Ds]$ by $\iota$. Summing over $\iota$ for fixed $L$ symmetrizes $\Theta_L$ over $\Aut(\Ds,q)$. This does not affect the result, since $E$ is $\Aut(\Ds,q)$-invariant, see~\eqref{eq:E-aut}. For independent replicas, each replica is an independent marked pair. Then $E(\tau)=\sum_xE_x(\tau)\ket{x}$ is a \emph{holomorphic} vector-valued Eisenstein series of weight $p/2$. Compare~\cite[(1.18)]{Bruinier2002}, which uses the dual Weil representation in weight $1+l/2$ for lattices of signature $(2,l)$, $(1,l-1)$ or $(0,l-2)$. Consequently $\cZ_{g,n;\chi}(\tau,\dots,\tau)$ is a component of a holomorphic vector-valued modular form of weight $np/2$ for $\rho^{\otimes n}$, and the generating function $\mathcal F_{g,\chi}$ of~\eqref{eq:egf} has mixed weight.

\subsection{Modular properties}
\label{sec:modular}
Let $\mathbf E(\btau):=E(\tau_1)\otimes\cdots\otimes E(\tau_n)\in\CC[\Ds]^{\otimes n}$. For $\bdelta=(\delta_i)\in\SL(2,\ZZ)^n$,
\begin{equation}
  \mathbf E(\bdelta\btau)=\prod_ij_{p,q}(\delta_i,\tau_i)\ \rho^{\otimes n}(\bdelta)\,\mathbf E(\btau)\;,
\end{equation}
and $\cZ_{g,n;\chi}=\cV_{g,n;\chi}(\mathbf E)$. More generally, for any $\Psi\in(\CC[\Ds]^{\otimes n})^*$ we set $\cZ[\Psi]:=\Psi(\mathbf E)$, which satisfies
\begin{equation}
  \cZ[\Psi](\bdelta\btau)=\prod_ij_{p,q}(\delta_i,\tau_i)\;\cZ\bigl[\Psi\circ\rho^{\otimes n}(\bdelta)\bigr](\btau)\;.
  \label{eq:modW}
\end{equation}
Thus $\{\cZ[\Psi]\}_\Psi$ is a finite-dimensional space of real-analytic functions on $\HH^n$ carrying the dual Weil representation, and the Seifert sums are the elements with $\Psi=\cV_{g,n;\chi}$.

Let $T_i$ and $S_i$ act on $\tau_i$ only. For $T_i$, the relation $E_x(\tau+1)=\e(q(x))E_x(\tau)$ and~\eqref{eq:main} give
\begin{equation}
  \cZ_{g,n;\chi}(T_i\btau)=\cZ_{g,n;\chi-1}(\btau)\;,
  \label{eq:Taction}
\end{equation}
so the shift $\tau_i\mapsto\tau_i+1$ is absorbed by the integer invariant $\chi$: the modular $T$-transformation is the Seifert move. For $S_i$, the relation $\rho(S)E(\tau)=\sum_\gamma j_{p,q}(\gamma,\tau)^{-1}\rho(\gamma S^{-1})^{-1}\ket0$ gives
\begin{equation}
  \cZ_{g,n;\chi}(S_i\btau)=j_{p,q}(S,\tau_i)\sum_{\bgamma}Z^{\rm pr}_{\rm CS}\bigl[g;\chi;\dots,\gamma_iS^{-1},\dots\bigr]\prod_kj_{p,q}(\gamma_k,\tau_k)^{-1}\;,
  \label{eq:Saction}
\end{equation}
where in~\eqref{eq:ZM} the $i$-th factor is $\bra x\rho(\gamma_iS^{-1})^{-1}\ket0$, evaluated on the matrix $\gamma_iS^{-1}$ itself. When its lower row $(-d_i,c_i)$ has $-d_i<0$, bringing it to the signed representative of Section~\ref{sec:review} multiplies the amplitude by $\e(\sigma/4)$ and the automorphy factor by the inverse phase.

The $S$-transformation relabels the slopes on $\mathbb T_i$ by $(c,d)\mapsto(-d,c)$, exchanging the roles of section and fiber. The \emph{set} of manifolds in the sum is unchanged, and only the assignment of automorphy factors changes. In terms of~\eqref{eq:modW}, however, $\cZ_{g,n;\chi}(S_i\btau)$ is $\cZ[\Psi]$ with $\Psi=\cV_{g,n;\chi}\circ\rho(S)_i$, which for $n\ge2$ is no longer diagonal in the fiber basis. Thus the Seifert functionals $\{\cV_{g,n;\chi}\}_\chi$ alone are not closed under $\SL(2,\ZZ)^n$: $T_i$ preserves them, but $S_i$ maps them to non-diagonal functionals. The Seifert--Eisenstein series is a component of the finite-dimensional modular multiplet $\{\cZ[\Psi]\}_\Psi$. In particular, it is modular under the congruence subgroup $\Gamma(N)^n$, where $N$ is the level of $(\Ds,q)$, i.e.\ the smallest positive integer with $Nq(x)\in\ZZ$ for all $x\in\Ds$. 

By~\eqref{eq:Taction}, shifting $\chi$ and shifting $\tau_i$ are the same operation. Then~\eqref{eq:main} gives $\cZ_{g,n;\chi+N}=\cZ_{g,n;\chi}$, and the finite Fourier transform
\begin{equation}
  \frac1N\sum_{\chi=0}^{N-1}\e\Bigl(\frac{\chi r}{N}\Bigr)\cZ_{g,n;\chi}(\btau)
  =|\Ds|^{\,g-1+\frac n2}\sum_{\substack{x\in\Ds\\ Nq(x)\equiv r\ (N)}}\ \prod_{i=1}^nE_x(\tau_i)
  \label{eq:chi-fourier}
\end{equation}
projects onto a fixed value $r/N$ of the topological spin $q(x)\bmod1$, without any convergence issue.\footnote{For $\Ds=\ZZ_3$, $N=3$, and this separates $E_0$ from $E_1$ in Section~\ref{sec:example}.} This periodicity is a property of the Chern--Simons weights. It does not identify manifolds with different $\chi$, and a sum over all integers $\chi$ still requires a separate weight. Similarly, the Fourier transform in $\chi$, $\sum_\chi\e(\chi\zeta)\cZ_{g,n;\chi}(\btau)$, is formally $|\Ds|^{g-1+n/2}\sum_x\delta_{\ZZ}\bigl(\zeta-q(x)\bigr)\prod_iE_x(\tau_i)$, where the variable $\zeta$ localizes onto a fixed value of the topological spin $q(x)\bmod1$. A convergent version uses a fugacity $y^\chi$ with $\chi\ge0$, which produces factors $(1-y\,\e(-q(x)))^{-1}$.

\subsection{Normalized Seifert data}
\label{sec:normalized}

Let us restrict~\eqref{eq:Zdef} to exceptional fillings, $c_i\geq2$ for all $i$, and write $\gamma_i=\bar\gamma_iT^{k_i}$, where $\bar\gamma_i$ has normalized lower row $(\alpha_i,\bar\beta_i)$ with $0<\bar\beta_i<\alpha_i$. By~\eqref{eq:seifert-move} the presentation $(\chi;\bgamma)$ describes the normalized manifold $\mathfrak s=(\chi';(\alpha_i,\bar\beta_i))$ with $\chi'=\chi+\sum_ik_i$. Using $j_{p,q}(\bar\gamma T^k,\tau)=j_{p,q}(\bar\gamma,\tau+k)$, we obtain
\begin{equation}
  \cZ^{\rm exc}_{g,n;\chi}(\btau)=\sum_{\mathfrak s\ \text{normalized, labeled}}Z^{\rm pr}_{\rm CS}[\mathfrak s]\;K_{\mathfrak s;\chi}(\btau),\qquad
  K_{\mathfrak s;\chi}(\btau):=\sum_{\substack{\bk\in\ZZ^n\\ \sum_ik_i=\chi'-\chi}}\ \prod_{i=1}^n j_{p,q}(\bar\gamma_i,\tau_i+k_i)^{-1}\;.
  \label{eq:normalized}
\end{equation}
Here $Z^{\rm pr}_{\rm CS}[\mathfrak s]$ is well defined because~\eqref{eq:ZM} is exactly invariant under Seifert moves. The kernel collects only the redundancy of the integer twists $\beta_i\mapsto\beta_i+k_i\alpha_i$ of the exceptional fillings. It contains no regular fillings, which were excluded from the start. Absorbing regular fillings as well would require keeping track of the removed legs, of the number of remaining exceptional fibers, and of the phase $\e(\sigma/8)$ per removed leg found below~\eqref{eq:ZMfr}. Each normalized Seifert datum $\mathfrak s$ thus comes with a partially periodized product of automorphy factors, summed over the $(n-1)$-dimensional sublattice $\sum k_i=\chi'-\chi$. After a Fourier transform in $\chi$ the constraint disappears, and the kernel factorizes into Lipschitz-type sums $\sum_{k\in\ZZ}\e(-k\zeta)\,j_{p,q}(\bar\gamma_i,\tau_i+k)^{-1}$, which converge for $p+q>2$.

\subsection{The generating function over fibers}
\label{sec:egf}
On the diagonal $\tau_1=\cdots=\tau_n=\tau$, the series
\[
  \cZ_{g,n;\chi}(\tau,\dots,\tau)=|\Ds|^{g-1+n/2}\sum_x\e(-\chi q(x))E_x(\tau)^n
\]
is a polynomial in the components of the vector-valued Eisenstein series. Let $\mathfrak s$ be an unordered collection of fillings with stabilizer $\Aut(\mathfrak s)=\prod_a\mathfrak S_{r_a}$, where $r_a$ counts the repeats of each filling. Since
\[
  \frac1{n!}\sum_{\text{labeled}}=\sum_{\text{unlabeled}}\frac{1}{|\Aut(\mathfrak s)|}\;,
\]
it is convenient to consider the exponential generating function
\begin{equation}
  \mathcal F_{g,\chi}(u;\tau):=\sum_{n\ge0}\frac{u^n}{n!}\cZ_{g,n;\chi}(\tau,\dots,\tau)
    =|\Ds|^{g-1}\sum_{x\in\Ds}\e(-\chi q(x))\,\exp\Bigl(u\,|\Ds|^{1/2}E_x(\tau)\Bigr)\;.
  \label{eq:egf}
\end{equation}
This describes a ``dilute gas'' of fibers on $\Sigma_g$. Splitting $E_x=\delta_{x,0}+R_x+\mathsf X_x$ according to $c=0$, $c=1$ and $c\ge2$, the exponential factorizes into three pieces: degenerate fillings, which project onto $x=0$; regular fibers, $R_x(\tau):=\e(\sigma/8)|\Ds|^{-1/2}\sum_{d}\e(-dq(x))\,j_{p,q}(\gamma_{1,d},\tau)^{-1}$, which renormalize $\chi$; and exceptional fibers $\mathsf X_x$.

These three pieces cannot be separated without losing modularity. Under $S$ acting on $\tau$, a filling $(c,d)$ is mapped to $(-d,c)$ (Section~\ref{sec:modular}). Thus the degenerate filling $(0,1)$ is exchanged with the regular fiber $(1,0)$, and a regular fiber $(1,d)$ with $|d|\ge2$ is mapped to the exceptional fiber $(|d|,\mp1)$. Hence $\delta_{x,0}$, $R_x$ and $\mathsf X_x$ are not separately modular covariant; only their sum $E_x$ is. Keeping only exceptional fibers, $\exp(u|\Ds|^{1/2}\mathsf X_x)$, would destroy the modular properties of Section~\ref{sec:modular}, so we keep the full generating function~\eqref{eq:egf}. Degenerate and regular fillings play the same role as the $c=0$ and $c=1$ terms of the lens-space sum~\cite{Ashwinkumar:2021kav}, which are likewise required by modular invariance.

If instead one wants each normalized Seifert datum with $n$ exceptional fibers to appear once with a single weight, the appropriate object is the normalized sum~\eqref{eq:normalized} over the exceptional part, in which the integer twists are absorbed into the kernels $K_{\mathfrak s;\chi}$, while regular and degenerate fillings are discarded. This is still not a sum over homeomorphism classes of the total space, since distinct normalized data can describe homeomorphic manifolds, for example lens spaces and other manifolds with several Seifert fibrations, or, if orientations are disregarded, data related by orientation reversal.

\subsection{One boundary: Dehn-filling Eisenstein series}
\label{sec:dehn}

The construction above uses $n$ boundary tori. With a single CFT torus there is an analogous construction for an arbitrary 3-manifold with torus boundary. Let $M_0$ be a compact oriented 3-manifold with $\partial M_0=T^2$ and an oriented basis $(\mathfrak l,\mathfrak m)$ of $H_1(\partial M_0)$. Let $\Psi_{M_0}\in\CC[\Ds]^*$ be its Chern--Simons functional, with some fixed choice of framing phase, and write $M_0(c,d)$ for the Dehn filling along $c\mathfrak l+d\mathfrak m$. As in~\eqref{eq:ZM}, we define the presentation-framed amplitude $Z^{\rm pr}_{\rm CS}[M_0;\gamma]:=\Psi_{M_0}(\rho(\gamma)^{-1}\ket0)\doteqf Z_{\rm CS}[M_0(c,d)]$ for $\gamma=\gamma_{c,d}$. The same argument as for~\eqref{eq:main} with $n=1$ gives, under the assumptions of~\eqref{eq:SW1},
\begin{equation}
  \vev{\Psi_{M_0}\bigl(\Theta(\tau;m)\bigr)}_{\cM_Q}=\Psi_{M_0}\bigl(E(\tau)\bigr)=\sum_{(c,d)}\frac{Z^{\rm pr}_{\rm CS}\bigl[M_0;\gamma_{c,d}\bigr]}{j_{p,q}(\gamma_{c,d},\tau)}\;.
  \label{eq:dehn}
\end{equation}
We call this the \emph{Dehn-filling Eisenstein series} of $M_0$. Under $\SL(2,\ZZ)$ the family $\{\Psi_{M_0}\circ\rho(\delta)\}$ corresponds to the same $M_0$ with a reparametrized boundary, so the modular group acts by permuting the Dehn fillings of a fixed $M_0$.

Two examples illustrate~\eqref{eq:dehn}. First, if $M_0$ is a solid torus with meridian $\mathfrak m$ and Wilson line $h$, then $\Psi=\bra h$ and we recover $E_h$ of~\cite{Ashwinkumar:2021kav}. The label $h$ is thus the Wilson line in the \emph{complementary} solid torus. Second, if $M_0$ is the complement of one fiber in a Seifert manifold, with the other $n-1$ fibers fixed, then
\begin{equation}
  \Psi_{M_0}(x)\doteqf|\Ds|^{g-1+n/2}\,\e(-\chi q(x))\prod_{i<n}\varepsilon_{Q,x}(\alpha_i,\beta_i)\;,
\end{equation}
and~\eqref{eq:dehn} sums over all fillings of the last boundary torus. The non-degenerate fillings give Seifert manifolds, while the degenerate filling gives a connected sum as in~\eqref{eq:degenerate}.

\paragraph{CFT interpretation of $\Psi(\Theta)$.}
Let $H\subset\Ds$ be a Lagrangian subgroup, i.e.\ $q|_H=0$ and $|H|^2=|\Ds|$. Then $\Psi_H:=\sum_{x\in H}\bra{x}$ gives $\Psi_H(\Theta)=\Theta_{\Lambda_H}$, the theta function of the even self-dual overlattice $\Lambda_H\supset\Lambda$. This is (up to $\eta$-factors) the partition function of a genuine Narain CFT, and $\Psi_H$ is the state of a topological boundary condition of the bulk TQFT~\cite{Kapustin:2010hk}. Note that $\bra0$ alone is of this form only if $\Ds$ is trivial. For a general $M_0$, $\Psi_{M_0}(\Theta)=\sum_x\Psi_{M_0}(x)\Theta_x$ is the torus partition function of the relative Narain theory of $\Lambda$ coupled to the bulk TQFT on $M_0$. In the language of SymTFT sandwiches~\cite{Freed:2022qnc}, the topological boundary of the sandwich is replaced by the 3-manifold $M_0$. The observable $\Psi_{M_0}(\Theta)$ is a linear combination of twisted-sector partition functions and is not modular invariant by itself. Its modular images $\Psi_{M_0}\circ\rho(\delta)$ correspond to reparametrizations of $\partial M_0$, so the family is closed under $\SL(2,\ZZ)$.

Equation~\eqref{eq:dehn} has the structure of ``TQFT gravity''~\cite{Dymarsky:2020qom,Aharony:2023zit,Barbar:2023ncl,Dymarsky:2024frx,Dymarsky:2025agh,Angelinos:2025zek,Barbar:2025vvf}, in which an ensemble average of a CFT observable equals a sum of TQFT amplitudes over a family of 3-manifolds. In those works the family consists mainly of handlebodies, which is the case where $M_0$ is a solid torus here, i.e.\ the lens-space sum of~\cite{Ashwinkumar:2021kav}. Non-handlebody topologies also appear there when the exponent of the discriminant group is not square-free~\cite{Dymarsky:2025agh,Angelinos:2025zek}. Equation~\eqref{eq:dehn} extends this to the family of all Dehn fillings of an arbitrary $M_0$. Here the CFT observable $\Psi_{M_0}(\Theta)$ specifies the fixed part $M_0$ of the topology, and the average generates the sum over its fillings.

\subsection{Example: \texorpdfstring{$\Ds=\ZZ_3$}{D=Z3}}
\label{sec:example}
The Gauss sums $\varepsilon_{Q,x}$ depend on $\Lambda$ only through the Weil representation, i.e.\ through the discriminant form $(\Ds,q)$ and $\sigma\bmod8$. As an example, take $\Lambda=A_2\oplus U\oplus U$, where $A_2$ is the root lattice with Gram matrix $\bigl(\begin{smallmatrix}2&-1\\-1&2\end{smallmatrix}\bigr)$ and $U$ is the even unimodular lattice of signature $(1,1)$ with Gram matrix $\bigl(\begin{smallmatrix}0&1\\1&0\end{smallmatrix}\bigr)$. Then $(p,q)=(4,2)$, so that $p+q=6>4$ and Siegel--Weil applies. Moreover, $\Ds=\ZZ_3$, $q(1)=q(2)=1/3$ and $\sigma=2$. Since $\varepsilon_{Q,-x}=\varepsilon_{Q,x}$, we have $E_1=E_2$, and~\eqref{eq:main} and~\eqref{eq:egf} become
\begin{equation}
  \begin{aligned}
    \cZ_{g,n;\chi}(\btau)       & =3^{\,g-1+\frac n2}\Bigl[\prod_{i=1}^nE_0(\tau_i)+2\,\e\bigl(-\tfrac{\chi}{3}\bigr)\prod_{i=1}^nE_1(\tau_i)\Bigr]\;, \\
    \mathcal F_{g,\chi}(u;\tau) & =3^{\,g-1}\Bigl[e^{\sqrt3\,uE_0(\tau)}+2\,\e\bigl(-\tfrac{\chi}{3}\bigr)e^{\sqrt3\,uE_1(\tau)}\Bigr]\;.
  \end{aligned}
\end{equation}
Only $\chi\bmod3$ enters, and $T$ acting on any $\tau_i$ cycles through the three values. The modular group acts on the pair $(E_0,E_1)$ through the two-dimensional subrepresentation of $\rho$ on $\{\ket0,\ket1+\ket2\}$.

The Fourier expansions are constrained by $E_x(\tau+1)=\e(q(x))E_x(\tau)$. The modes of $E_0$ lie in $\ZZ$ and those of $E_1$ in $\tfrac13+\ZZ$, so only $E_0$ has a constant term. To compute it, take $\Lambda=A_2$ in~\eqref{eq:eps}; it has the same discriminant form and the same $\sigma$, hence the same $\varepsilon_{Q,x}$. For $3\nmid c$ the sum in~\eqref{eq:eps} is the Gauss sum of the binary form $\ell_1^2-\ell_1\ell_2+\ell_2^2$ of discriminant $-3$, which equals $c\,\chi_{-3}(c)$ independently of $d$, where $\chi_{-3}$ is the non-trivial character mod $3$. Hence $\varepsilon_{Q,0}(c,d)=i\chi_{-3}(c)/\sqrt3$. For $3\mid c$ one finds instead $\varepsilon_{Q,0}(c,d)=\chi_{-3}(d)$, so the sum over $d$ vanishes. Poisson summation over $d$, with $\int_{\RR}(t+iy)^{-2}(t-iy)^{-1}\,dt=-i\pi/(2y^2)$ and $\sum_c\chi_{-3}(c)\varphi(c)c^{-s}=L(s-1,\chi_{-3})/L(s,\chi_{-3})$, then gives
\begin{equation}
  E_0(\tau)=1+\frac{81\,L(2,\chi_{-3})}{8\pi^2}\,\frac{1}{y^2}+\bigl(\text{non-zero modes}\bigr),\qquad y:=\operatorname{Im}\tau\;,
  \label{eq:E0const}
\end{equation}
where we used $L(3,\chi_{-3})=4\pi^3/(81\sqrt3)$. Numerically, we have $81L(2,\chi_{-3})/(8\pi^2)\simeq0.8015$. Consequently, the term of $\cZ_{g,n;\chi}(\btau)$ that is constant in every $\tau_i$ is
\begin{equation}
  3^{\,g-1+\frac n2}\prod_{i=1}^n\Bigl(1+\frac{81\,L(2,\chi_{-3})}{8\pi^2\,y_i^2}\Bigr)\;,
\end{equation}
which is independent of $\chi$, since the dependence on $\chi$ enters only through the modes of $E_1$, which carry topological spin $q(1)=1/3$. The $y^{-2}$ terms are the contributions of the solid tori with $c\ge1$ to the zero modes.

Table~\ref{tab:example} lists the canonically framed partition functions of some Seifert manifolds, computed from the surgery formula~\eqref{eq:surgery}. We find that they agree with~\eqref{eq:ZMcan} in each case. The magnitudes follow the pattern of~\eqref{eq:linking}. In these examples $|Z^{\rm can}_{\rm CS}|=3^{(b_1-1)/2}\,|H_1(M)_{\rm tors}\otimes\ZZ_3|^{1/2}$. In particular, manifolds with $b_1=0$ and without $3$-torsion, such as the Poincar\'e sphere or the prism manifold, all give $3^{-1/2}$.

\begin{table}[h]
  \centering
  \setlength{\abovecaptionskip}{0pt}
  \setlength{\belowcaptionskip}{8pt}
  \caption{Canonically framed Abelian Chern--Simons invariants for $\Ds=\ZZ_3$, $\sigma=2$, with the orientation of Section~\ref{sec:seifert}. By the convention of Section~\ref{sec:dict}, the presentation-framed amplitude $Z^{\rm pr}_{\rm CS}$ of~\eqref{eq:ZM} is the complex conjugate of the invariant tabulated here, up to the phase in~\eqref{eq:ZMcan}.}
  \label{tab:example}
  \small
  \begin{tabular}{@{}llll@{}}
    \toprule
    $M(g;\chi;(\alpha_i,\beta_i))$ &                   & $H_1(M)$           & $Z^{\rm can}_{\rm CS}[M]$ \\
    \midrule
    $M(0;1;\,\emptyset)$           & $S^3$             & $0$                & $1/\sqrt3$                \\
    $M(0;3;\,\emptyset)$           & $L(3,1)$          & $\ZZ_3$            & $-i$                      \\
    $M(0;0;(3,1),(3,1))$           & lens space        & $\ZZ_6$            & $i$                       \\
    $M(0;0;(3,1),(3,-1))$          & $S^2\times S^1$   & $\ZZ$              & $1$                       \\
    $M(0;-1;(2,1),(3,1),(5,1))$    & Poincar\'e sphere & $0$                & $1/\sqrt3$                \\
    $M(0;0;(2,1),(2,1),(2,-1))$    & prism manifold    & $\ZZ_2^2$          & $1/\sqrt3$                \\
    $M(0;0;(3,1),(3,1),(3,1))$     &                   & $\ZZ_3\oplus\ZZ_9$ & $-\sqrt3$                 \\
    $M(0;-1;(3,1),(3,1),(3,1))$    & ($\mathsf e=0$)   & $\ZZ\oplus\ZZ_3$   & $-i\sqrt3$                \\
    $M(1;0;\,\emptyset)$           & $T^3$             & $\ZZ^3$            & $3$                       \\
    $M(1;1;(3,1))$                 &                   & $\ZZ^2\oplus\ZZ_4$ & $\sqrt3$                  \\
    \bottomrule
  \end{tabular}
\end{table}

\section{Common modulus: wormholes and self-gluings}
\label{sec:common}
We now let all interfaces carry the \emph{same} Narain modulus $m$. Then $\cO_{g,n;\chi}(\btau;m,\dots,m)$ equals $\cV_{g,n;\chi}\bigl(\Theta^{[n]}(\diag\btau;m)\bigr)$, and the degree-$n$ Siegel--Weil formula (Section~\ref{sec:SWn}) gives
\begin{equation}
  \vev{\cO_{g,n;\chi}(\btau;m,\dots,m)}_{\cM_Q}=\cV_{g,n;\chi}\bigl(E^{[n]}(\diag\btau)\bigr)\;.
  \label{eq:common}
\end{equation}
Throughout this section we assume $p+q>2n+2$, so that $E^{[n]}$ converges absolutely and the degree-$n$ Siegel--Weil formula holds without regularization~\cite{Weil1965}; what happens for smaller $p+q$ is discussed at the end of Section~\ref{sec:SWn}. For $pq\neq0$, the convergence of the common-modulus average at fixed $(p,q)$ is guaranteed by Weil's criterion only for finitely many $n$. Therefore, unlike the replica average~\eqref{eq:main}, we do not sum it over all $n$ as in~\eqref{eq:egf}. In this section we relate~\eqref{eq:common} to the Seifert sum~\eqref{eq:main}.

\subsection{Degree-\texorpdfstring{$n$}{n} Siegel--Weil and the cumulant expansion}
\label{sec:SWn}
For $\Omega$ in the Siegel upper half space $\HH_n$, let
\[
  \Theta^{[n]}(\Omega;m):=\sum_{\bx\in\Ds^n}\Theta^{[n]}_{\bx}(\Omega;m)\ket{\bx}\in\CC[\Ds]^{\otimes n}
\]
be the degree-$n$ (or genus-$n$) Siegel--Narain theta vector. It transforms under $\Sp(2n,\ZZ)$ through the Weil representation $\rho_n$, with automorphy factor $j^{[n]}_{p,q}(\gamma,\Omega):=\det(C\Omega+D)^{p/2}\det(C\bar\Omega+D)^{q/2}$, where $C$ and $D$ are the lower $n\times n$ blocks of a $(2n)\times (2n)$-matrix $\gamma=\bigl(\begin{smallmatrix}A&B\\C&D\end{smallmatrix}\bigr)$. The Siegel--Eisenstein vector is
\begin{equation}
  E^{[n]}(\Omega):=\sum_{\gamma\in P_n\backslash\Sp(2n,\ZZ)}j^{[n]}_{p,q}(\gamma,\Omega)^{-1}\rho_n(\gamma)^{-1}\ket{0}^{\otimes n}\;,
\end{equation}
where $P_n$ is the Siegel parabolic subgroup ($C=0$). Under the convergence assumptions of~\cite{Weil1965,Ashwinkumar:2021kav} (for instance $p+q>2n+2$) one has $\vev{\Theta^{[n]}(\Omega;m)}_{\cM_Q}=E^{[n]}(\Omega)$~\cite{Siegel1944,Weil1965,Ashwinkumar:2021kav,Maloney:2020nni}. For the extension to smaller $p+q$ see the end of this subsection. The cosets $P_n\backslash\Sp(2n,\ZZ)$ are in bijection with primitive Lagrangian sublattices $L_\gamma\subset\ZZ^{2n}$ (the row space of $(C\ D)$).

On the block-diagonal locus $\Omega=\diag(\tau_1,\dots,\tau_n)$ we have $\Theta^{[n]}=\bigotimes_i\Theta(\tau_i;m)$ at a \emph{common} modulus $m$. Moreover $\rho_n|_{\SL(2,\ZZ)^n}=\rho^{\otimes n}$ and $j^{[n]}_{p,q}(\gamma_1\oplus\cdots\oplus\gamma_n,\diag\btau)=\prod_ij_{p,q}(\gamma_i,\tau_i)$.

Write $\ZZ^{2n}=\bigoplus_{i=1}^nU_i$, with $U_i\cong\ZZ^2$ the homology of $\mathbb T_i$, and for a subset $B\subset\{1,\dots,n\}$ let $U_B:=\bigoplus_{i\in B}U_i$. We say that a Lagrangian $L\subset\ZZ^{2n}$ is \emph{subordinate} to a set partition $\pi$ of $\{1,\dots,n\}$ if $L=\bigoplus_{B\in\pi}(L\cap U_B)$, and \emph{indecomposable} if it is subordinate only to the trivial partition. Every primitive Lagrangian $L$ is subordinate to a unique finest partition $\pi(L)$, and each block $L\cap U_B$, $B\in\pi(L)$, is an indecomposable primitive Lagrangian in $U_B$. To see this, suppose that $L$ is subordinate to both $\pi$ and $\pi'$. Take $v\in L\cap U_B$ with $B\in\pi$, and decompose $v=\sum_{B'\in\pi'}v_{B'}$ with $v_{B'}\in L\cap U_{B'}$. Then $v_{B'}$ is the projection of $v$ to $U_{B'}$, so it lies in $U_B\cap U_{B'}=U_{B\cap B'}$. Hence $L$ is subordinate to the common refinement $\pi\wedge\pi'$, and a finest partition exists. Each block $L\cap U_B$ is isotropic. Its rank is $|B|$, because the ranks add up to $n$ and each block is at most half-dimensional. It is primitive because $L$ is.

Let $E^{[B],\rm conn}(\btau_B)$ denote the part of the sum $E^{[|B|]}(\diag\btau_B)$ that runs over indecomposable Lagrangians only,
\[
  E^{[B],\rm conn}(\btau_B):=\sum_{\substack{\gamma\in P_{|B|}\backslash\Sp(2|B|,\ZZ)\\ L_\gamma\ \text{indecomposable}}}j^{[|B|]}_{p,q}(\gamma,\diag\btau_B)^{-1}\rho_{|B|}(\gamma)^{-1}\ket{0}^{\otimes|B|}\;.
\]
Since $|B|\le n$, the assumption $p+q>2n+2$ guarantees absolute convergence and degree-$|B|$ Siegel--Weil for every $B$, and we have
\begin{equation}
  E^{[n]}(\diag\btau)=\sum_{\pi}\ \bigotimes_{B\in\pi}E^{[B],\rm conn}(\btau_B),\qquad
  E^{[B],\rm conn}(\btau_B)=\Bigl\langle\bigotimes_{i\in B}\Theta(\tau_i;m)\Bigr\rangle_{\rm c}\;,
  \label{eq:cumulant}
\end{equation}
where $\vev{\cdots}_{\rm c}$ is the joint cumulant with respect to the average over $\cM_Q$ at the common modulus $m$. It is defined recursively by the moment--cumulant relation: the moment $\vev{\bigotimes_{i\in B}\Theta(\tau_i;m)}_{\cM_Q}$ is the sum, over the set partitions of $B$, of the tensor products of the cumulants of the blocks. For instance, $\vev{\Theta(\tau_1;m)}_{\rm c}=E(\tau_1)$ and
\[
  \vev{\Theta(\tau_1;m)\otimes\Theta(\tau_2;m)}_{\rm c}=\int_{\cM_Q}dm\,\Theta(\tau_1;m)\otimes\Theta(\tau_2;m)-E(\tau_1)\otimes E(\tau_2)\;.
\]
Correspondingly, $E^{[\{1\}],\rm conn}(\tau_1)=E(\tau_1)$ and
\[
  E^{[\{1,2\}],\rm conn}(\tau_1,\tau_2)=E^{[2]}(\diag(\tau_1,\tau_2))-E(\tau_1)\otimes E(\tau_2)\;.
\]
Thus the decomposition of the Siegel--Eisenstein series by indecomposable Lagrangians coincides with the cumulant expansion. To see this, we split the coset sum according to $\pi(L)$. For fixed $\pi$, each block $L\cap U_B$ is the row space of $(C_B\ D_B)$ for some $\gamma_B\in\Sp(2|B|,\ZZ)$, and we can take $\gamma_L$ to be the symplectic direct sum of the $\gamma_B$. Since $\rho_n$ restricts to the tensor product on $\prod_B\Sp(2|B|,\ZZ)$, and $C\Omega+D$ is block diagonal for $\Omega=\diag\btau$, both the Weil coefficient and the automorphy factor factorize over blocks, which gives the first equation. Applying the first equation to each subset $B$ and using degree-$|B|$ Siegel--Weil gives
\[
  \Bigl\langle\bigotimes_{i\in B}\Theta(\tau_i;m)\Bigr\rangle_{\cM_Q}
  =\sum_{\pi'}\ \bigotimes_{B'\in\pi'}E^{[B'],\rm conn}(\btau_{B'})\;,
\]
where $\pi'$ runs over the set partitions of $B$. This is the moment--cumulant relation, which determines the cumulants uniquely by induction on $|B|$, starting from~\eqref{eq:SW1} for $|B|=1$. Hence $E^{[B],\rm conn}$ is the joint cumulant, which is the second equation. In particular, the all-singletons term in~\eqref{eq:cumulant} is $\bigotimes_iE(\tau_i)$, the replica average~\eqref{eq:main}. The organization of the Siegel--Eisenstein series by Lagrangian sublattices, with decomposable (block-diagonal) sublattices giving the factorized contribution of disconnected geometries and indecomposable ones the connected correlator, including the $T^2\times I$ wormhole, was discussed in~\cite[Section~3.2]{Maloney:2020nni} and~\cite[Sections~2,~4]{Collier:2021rsn}. Here we add the identification with cumulants and, below, with self-gluings of the core.

Equation~\eqref{eq:cumulant} is the analogue of~\cite[(3.15)]{Marolf:2020xie} in the two-dimensional topological gravity of Marolf and Maxfield. There, $\vev{Z^n}/\vev{1}=\sum_\pi\lambda^{|\pi|}$ is a sum over partitions of the boundaries into connected components of spacetime, and each component contributes the genus-summed amplitude $\lambda$ (Section~\ref{sec:MM}). Here each block contributes the sum $E^{[B],\rm conn}$ over indecomposable Lagrangians. For $n=2$ the two terms of~\cite{Marolf:2020xie}, a pair of discs and a cylinder, correspond to $E\otimes E$ (pairs of solid tori) and to $E^{[2],\rm conn}$ (the wormholes of Section~\ref{sec:selfglue}). After reduction along the fiber, a solid torus $V_{(c,d)}$ is a disc with a cone point of order $c$, and a fiber-preserving wormhole is an annulus with at most one cone point.

The assumption $p+q>2n+2$ is needed only for the Lagrangian side of~\eqref{eq:cumulant}. The theta integrals defining the moments converge under Weil's weaker condition that $Q$ is anisotropic over $\QQ$ or $p+q-r>|B|+1$, where $r$ is the Witt index of $Q$, i.e.\ the dimension of a maximal isotropic subspace of $\Lambda\otimes\QQ$~\cite{Weil1965}. In this range the cumulants $\vev{\bigotimes_{i\in B}\Theta(\tau_i;m)}_{\rm c}$ are well defined and finite, and the moments are given by the Siegel--Weil formula in its extended form, with $E^{[|B|]}$ defined by analytic continuation in the Eisenstein parameter~\cite{KudlaRallis1988,KudlaRallis1988II}; see also~\cite{Kudla2008}. (If $Q$ is anisotropic and $p+q\le|B|+1$, the moment is $\tfrac12E^{[|B|]}$.) For $p+q\le2|B|+2$, however, the coset sum defining $E^{[|B|]}$ is not absolutely convergent, so it cannot be split termwise by $\pi(L)$. In this range~\eqref{eq:cumulant} should be read as a definition of $E^{[B],\rm conn}$ by the cumulants. Its expansion over indecomposable Lagrangians, and hence the wormhole interpretation below, is then only formal. Beyond Weil's range the convergence of the averages is no longer guaranteed, and we do not consider this case.

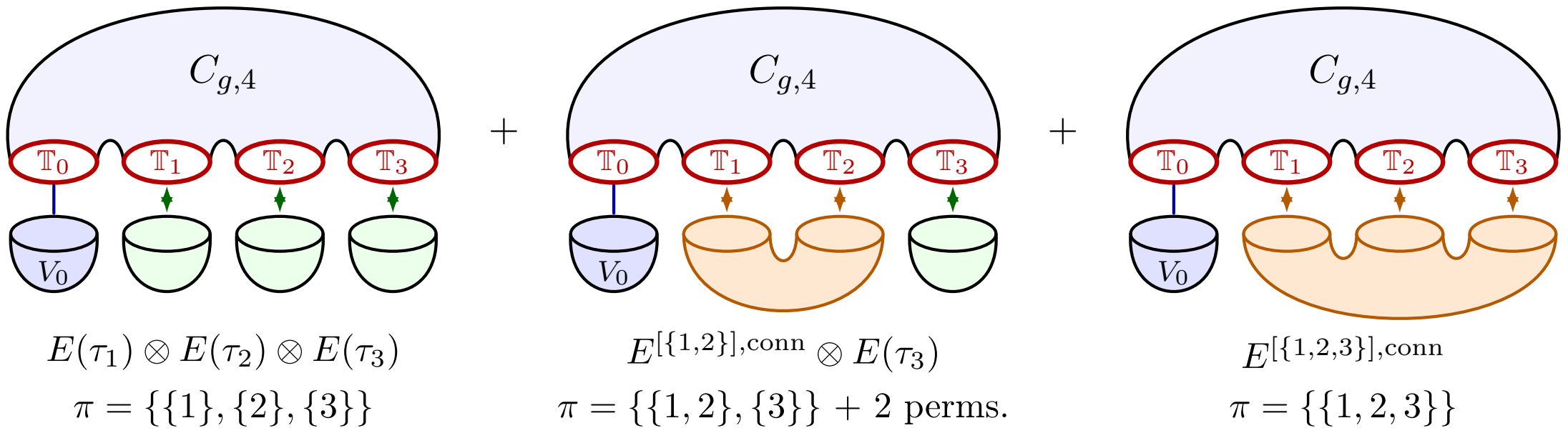
\begin{figure}[t]
  \centering
  \begin{tikzpicture}[line cap=round,line join=round,thick,scale=0.95,every node/.style={transform shape}]
    \def\connectedfill{orange!18}
    \def\connecteddraw{orange!70!black}
    \def\solidtorus#1{%
      \draw[{Latex[length=1.5mm]}-{Latex[length=1.5mm]},green!40!black] (#1,-0.23) -- (#1,-0.5);
      \fill[green!8] (#1-0.42,-0.7) .. controls (#1-0.47,-1.45) and (#1+0.47,-1.45) .. (#1+0.42,-0.7) -- cycle;
      \draw[fill=green!8] (#1,-0.7) ellipse (0.42 and 0.17);
      \draw (#1-0.42,-0.7) .. controls (#1-0.47,-1.45) and (#1+0.47,-1.45) .. (#1+0.42,-0.7);
    }
    \def\leg#1{%
      \draw[{Latex[length=1.5mm]}-{Latex[length=1.5mm]},\connecteddraw] (#1,-0.23) -- (#1,-0.5);
      \draw[\connecteddraw,fill=\connectedfill] (#1,-0.7) ellipse (0.42 and 0.17);
    }
    \def\core{%
      \fill[blue!5] (-2.07,0) .. controls (-2.3,1.0) and (-1.2,1.5) .. (0,1.5)
      .. controls (1.2,1.5) and (2.3,1.0) .. (2.07,0) -- (1.23,0)
      .. controls (1.2,0.28) and (1.0,0.28) .. (0.97,0) -- (0.13,0)
      .. controls (0.1,0.28) and (-0.1,0.28) .. (-0.13,0) -- (-0.97,0)
      .. controls (-1.0,0.28) and (-1.2,0.28) .. (-1.23,0) -- cycle;
      \draw (-2.07,0) .. controls (-2.3,1.0) and (-1.2,1.5) .. (0,1.5)
      .. controls (1.2,1.5) and (2.3,1.0) .. (2.07,0);
      \draw (1.23,0) .. controls (1.2,0.28) and (1.0,0.28) .. (0.97,0);
      \draw (0.13,0) .. controls (0.1,0.28) and (-0.1,0.28) .. (-0.13,0);
      \draw (-0.97,0) .. controls (-1.0,0.28) and (-1.2,0.28) .. (-1.23,0);
      \node at (0,0.85) {$\Mcore_{g,4}$};
      \foreach \x/\lab in {-1.65/0,-0.55/1,0.55/2,1.65/3} {
          \draw[very thick,red!70!black,fill=white] (\x,0) ellipse (0.42 and 0.2);
          \node[red!70!black,font=\scriptsize] at (\x,0) {$\mathbb T_\lab$};
        }
      \draw[blue!60!black] (-1.65,-0.23) -- (-1.65,-0.5);
      \fill[blue!12] (-2.07,-0.7) .. controls (-2.12,-1.45) and (-1.18,-1.45) .. (-1.23,-0.7) -- cycle;
      \draw[fill=blue!12] (-1.65,-0.7) ellipse (0.42 and 0.17);
      \draw (-2.07,-0.7) .. controls (-2.12,-1.45) and (-1.18,-1.45) .. (-1.23,-0.7);
      \node[font=\scriptsize] at (-1.65,-1.08) {$V_0$};
    }
    \begin{scope}[xshift=-5.45cm]
      \core
      \solidtorus{-0.55}
      \solidtorus{0.55}
      \solidtorus{1.65}
      \node[font=\small] at (0,-1.85) {$E(\tau_1)\otimes E(\tau_2)\otimes E(\tau_3)$};
      \node[font=\small] at (0,-2.4) {$\pi=\{\{1\},\{2\},\{3\}\}$};
    \end{scope}
    \node at (-2.72,0.3) {$+$};
    \begin{scope}
      \core
      \fill[\connectedfill] (-0.97,-0.7) .. controls (-0.97,-1.7) and (0.97,-1.7) .. (0.97,-0.7)
      -- (0.13,-0.7) .. controls (0.13,-1.05) and (-0.13,-1.05) .. (-0.13,-0.7) -- cycle;
      \draw[\connecteddraw] (-0.97,-0.7) .. controls (-0.97,-1.7) and (0.97,-1.7) .. (0.97,-0.7);
      \draw[\connecteddraw] (0.13,-0.7) .. controls (0.13,-1.05) and (-0.13,-1.05) .. (-0.13,-0.7);
      \leg{-0.55}
      \leg{0.55}
      \solidtorus{1.65}
      \node[font=\small] at (0,-1.85) {$E^{[\{1,2\}],\rm conn}\otimes E(\tau_3)$};
      \node[font=\small] at (0,-2.4) {$\pi=\{\{1,2\},\{3\}\}$ $+$ 2 perms.};
    \end{scope}
    \node at (2.72,0.3) {$+$};
    \begin{scope}[xshift=5.45cm]
      \core
      \fill[\connectedfill] (-0.97,-0.7) .. controls (-0.97,-1.8) and (2.07,-1.8) .. (2.07,-0.7)
      -- (1.23,-0.7) .. controls (1.23,-1.02) and (0.97,-1.02) .. (0.97,-0.7)
      -- (0.13,-0.7) .. controls (0.13,-1.02) and (-0.13,-1.02) .. (-0.13,-0.7) -- cycle;
      \draw[\connecteddraw] (-0.97,-0.7) .. controls (-0.97,-1.8) and (2.07,-1.8) .. (2.07,-0.7);
      \draw[\connecteddraw] (1.23,-0.7) .. controls (1.23,-1.02) and (0.97,-1.02) .. (0.97,-0.7);
      \draw[\connecteddraw] (0.13,-0.7) .. controls (0.13,-1.02) and (-0.13,-1.02) .. (-0.13,-0.7);
      \leg{-0.55}
      \leg{0.55}
      \leg{1.65}
      \node[font=\small] at (0,-1.85) {$E^{[\{1,2,3\}],\rm conn}$};
      \node[font=\small] at (0,-2.4) {$\pi=\{\{1,2,3\}\}$};
    \end{scope}
  \end{tikzpicture}
  \caption{The cumulant expansion~\eqref{eq:common-decomp} for $n=3$, assuming $p+q>8$. The core $\Mcore_{g,4}$ has four boundary tori: $\mathbb T_0$ is filled by the fixed solid torus $V_0$, and the Narain theory with a common modulus $m$ lives on $\mathbb T_1,\mathbb T_2,\mathbb T_3$. Each of the five set partitions $\pi$ of $\{1,2,3\}$ gives one term. Left: all blocks are singletons, which gives the Seifert sum over independent solid tori (green). Middle: the block $\{1,2\}$ is the sum over wormholes gluing the core to itself (Section~\ref{sec:selfglue}), and likewise for $\{1,3\}$ and $\{2,3\}$. Right: the single block is the sum over indecomposable Lagrangians of $U_1\oplus U_2\oplus U_3$, drawn schematically as a connected piece with three boundary tori. Cobordisms realizing the indecomposable Lagrangians are constructed in Section~\ref{sec:selfglue} only for blocks with $|B|=2$.}
  \label{fig:cumulant}
\end{figure}

Contracting~\eqref{eq:cumulant} with~\eqref{eq:core}, the common-modulus average becomes
\begin{equation}
  \begin{aligned}
    \vev{\cO_{g,n;\chi}(\btau;m,\dots,m)}_{\cM_Q} & =\underbrace{\cZ_{g,n;\chi}(\btau)}_{\text{Seifert sum}}                                                                                       \\
                                                  & \quad+|\Ds|^{g-1+\frac n2}\sum_{x}\e(-\chi q(x))\sum_{\pi\neq\{\{1\},\dots,\{n\}\}}\ \prod_{B\in\pi}E^{[B],\rm conn}_{(x,\dots,x)}(\btau_B)\;.
  \end{aligned}
  \label{eq:common-decomp}
\end{equation}
Figure~\ref{fig:cumulant} illustrates this decomposition for $n=3$, where the core $\Mcore_{g,4}$ has $n+1=4$ boundary tori, one of which is filled by $V_0$.

\subsection{Two boundaries and self-gluings}
\label{sec:selfglue}
For $n=2$, an indecomposable $L\subset U_1\oplus U_2$ meets neither $U_1$ nor $U_2$. It is therefore the graph $L_A:=\{(v,Av)\}$ of some $A\in\GL(2,\QQ)$ with $\det A=-1$, restricted to $v\in\ZZ^2\cap A^{-1}\ZZ^2$. When $A\in\GL(2,\ZZ)$, $L_A$ is the kernel of $H_1(\partial X)\to H_1(X)$ for the mapping cylinder $X\cong T^2\times I$ of the orientation-reversing identification $\varphi=-A:\mathbb T_1\to\mathbb T_2$ (a cycle $c_1$ is homologous in $X$ to $\varphi(c_1)$, so $(c_1,-\varphi(c_1))$ lies in the kernel). For a coset representative $\gamma_A$ with $L_{\gamma_A}=L_A$, the coefficient $\bra{x_1,x_2}\rho_2(\gamma_A)^{-1}\ket{0,0}$ is then $|\Ds|^{-1/2}$ times the amplitude of this mapping cylinder, up to a phase (see below~\eqref{eq:Aalphabeta}). The factor $|\Ds|^{-1/2}$ is already visible for $A=\diag(1,-1)$: the identity cylinder has the state $\sum_x\ket{x,x}$ of norm $|\Ds|^{1/2}$, whereas the unitary image $\rho_2(\gamma_A)^{-1}\ket{0,0}$ has norm one. This is the torus wormhole of~\cite{Maloney:2020nni,Cotler:2020ugk,Collier:2021rsn} (see also~\cite{Kawamoto:2026jdz}), whose automorphy factor is built from $\det(C\diag(\tau_1,\tau_2)+D)$, of bidegree $(1,1)$ in $(\tau_1,\tau_2)$. Contracting with $\cV_{g,2;\chi}$ glues $\mathbb T_1$ to $\mathbb T_2$ through this cylinder.

This interpretation can be made precise, up to normalization and phase, for all indecomposable $L\subset U_1\oplus U_2$, integral or rational, in the following sense. For each $L$ we construct a cobordism $X$ whose Chern--Simons state is the corresponding term. The Lagrangian does not determine $X$ up to homeomorphism: for instance, a connected sum of $X$ with an integral homology sphere in its interior leaves the kernel below, and the Abelian Chern--Simons state, unchanged. The argument uses two facts.

First, for a primitive Lagrangian $L\subset\ZZ^{2n}$, consider the (Weyl-ordered) Wilson-line operators along the cycles in $L$, with charges in $\Ds$. They generate a Lagrangian subgroup $L\otimes\Ds$ of the finite Heisenberg group acting on $\CC[\Ds]^{\otimes n}$. Their joint eigenspaces are therefore one-dimensional, and $\rho_n(\gamma_L)^{-1}\ket0^{\otimes n}$ spans the one on which they act trivially. (For $L$ the row space of $(0\ \mathbbm 1)$, this is the statement that the joint eigenvectors of the holonomies are the basis vectors; the general case follows by $\Sp(2n,\ZZ)$ covariance.)

Second, let $X$ be a 3-manifold with $\partial X=\mathbb T_1\sqcup\mathbb T_2$, glued to the core along these tori, and suppose that the kernel of $H_1(\partial X;\ZZ)\to H_1(X;\ZZ)$ is $L$ in the bases of the core. Then Wilson lines along cycles of $L$ bound surfaces in $X$, so these operators act trivially on $Z_{\rm CS}[X]$. Hence $Z_{\rm CS}[X]$ is proportional to $\rho_2(\gamma_L)^{-1}\ket{0,0}$.

Such an $X$ exists for every indecomposable $L_A$. Let $X_{\alpha,\beta}:=(\Sigma_{0,3}\times S^1)\cup V$ be the annulus times $S^1$ with one exceptional fiber $(\alpha,\beta)$, i.e.\ with the third boundary filled along $\alpha\mathfrak l_3+\beta\mathfrak m$. Its first homology $H_1(X_{\alpha,\beta})=\langle\mathfrak l_1,\mathfrak l_3,\mathfrak m\mid\alpha\mathfrak l_3+\beta\mathfrak m=0\rangle\cong\ZZ^2$ is torsion-free. A short computation shows that the kernel is the graph of
\begin{equation}
  A_{\alpha,\beta}:=\begin{pmatrix}1&0\\-\beta/\alpha&-1\end{pmatrix}\qquad\text{in the bases }(\mathfrak l_i,\mathfrak m_i)\text{ of }\partial X_{\alpha,\beta}\;.
  \label{eq:Aalphabeta}
\end{equation}
We glue $X_{\alpha,\beta}$ to the core so that the bases glue orientably, which identifies $\mathfrak l\mapsto-\mathfrak l'$ and $\mathfrak m\mapsto\mathfrak m'$, a map of determinant $-1$. In the bases of the core, the kernel then becomes the graph of $\diag(-1,1)A_{\alpha,\beta}\diag(-1,1)=A_{\alpha,-\beta}$. By~\eqref{eq:core}, the Chern--Simons state of $X_{\alpha,\beta}$ on the boundaries of the core is, in the fiber basis, $Z_{\rm CS}[X_{\alpha,\beta}]\doteqf|\Ds|^{1/2}\sum_x\varepsilon_{Q,x}(\alpha,\beta)\ket{x,x}$. Using the quasi-periodicity $\varepsilon_{Q,x+\alpha y}(\alpha,\beta)=\e\bigl(-\beta(x,y)-\alpha\beta q(y)\bigr)\varepsilon_{Q,x}(\alpha,\beta)$, which follows from~\eqref{eq:eps}, one checks directly that the Weyl-ordered operators of $L_{A_{\alpha,-\beta}}$ act trivially on this vector. Since $\rho_2$ is unitary, $\rho_2(\gamma_L)^{-1}\ket{0,0}$ has unit norm, while $Z_{\rm CS}[X_{\alpha,\beta}]$ has norm $|\Ds|^{1/2}\bigl(\sum_x|\varepsilon_{Q,x}|^2\bigr)^{1/2}=|\Ds|^{1/2}$. Thus the term with Lagrangian $L_{A_{\alpha,-\beta}}$ is exactly $|\Ds|^{-1/2}Z_{\rm CS}[X_{\alpha,\beta}]$ up to a phase. This phase is the analogue of the framing phase in~\eqref{eq:ZMcan}. It depends on the choice of the coset representative $\gamma_L$ (the dependence is compensated by the automorphy factor in $E^{[2]}$), and it does not affect the identification of the topology. For $\alpha=1$, $X_{1,\beta}$ is the mapping cylinder twisted by $\beta$, which shifts $\chi$, and hence $\mathsf e$, by $\beta$.

A general rational $A$ with $\det A=-1$ has elementary divisors $(1/k,k)$ with $k\in\ZZ_{>0}$, and $A_{k,\beta}$ has the same elementary divisors for every $\beta$ coprime to $k$. Hence $A=\delta_2A_{k,\beta}\delta_1$ with $\delta_{1,2}\in\GL(2,\ZZ)$. Here $\det\delta_1=\det\delta_2$, and we may take both equal to $-1$: otherwise we replace $(\delta_1,\beta,\delta_2)$ by $(\diag(-1,1)\delta_1,-\beta,\delta_2\diag(-1,1))$. Then $L_A$ is realized by $X_{k,\beta}$ glued to the core through the orientation-reversing maps $\delta_1$ and $\delta_2^{-1}$. While $k$ is fixed by $A$, the choice of $\beta$ is part of the realization. The gluing above is the case $\delta_1=\delta_2=\diag(-1,1)$. In this sense the rational Lagrangians are realized by wormholes with one exceptional fiber of order $k$, the ``denominator'' of $A$; the integral ones ($k=1$) are realized by the torus wormholes above. The closed manifolds discussed below are those obtained from this choice of realization.

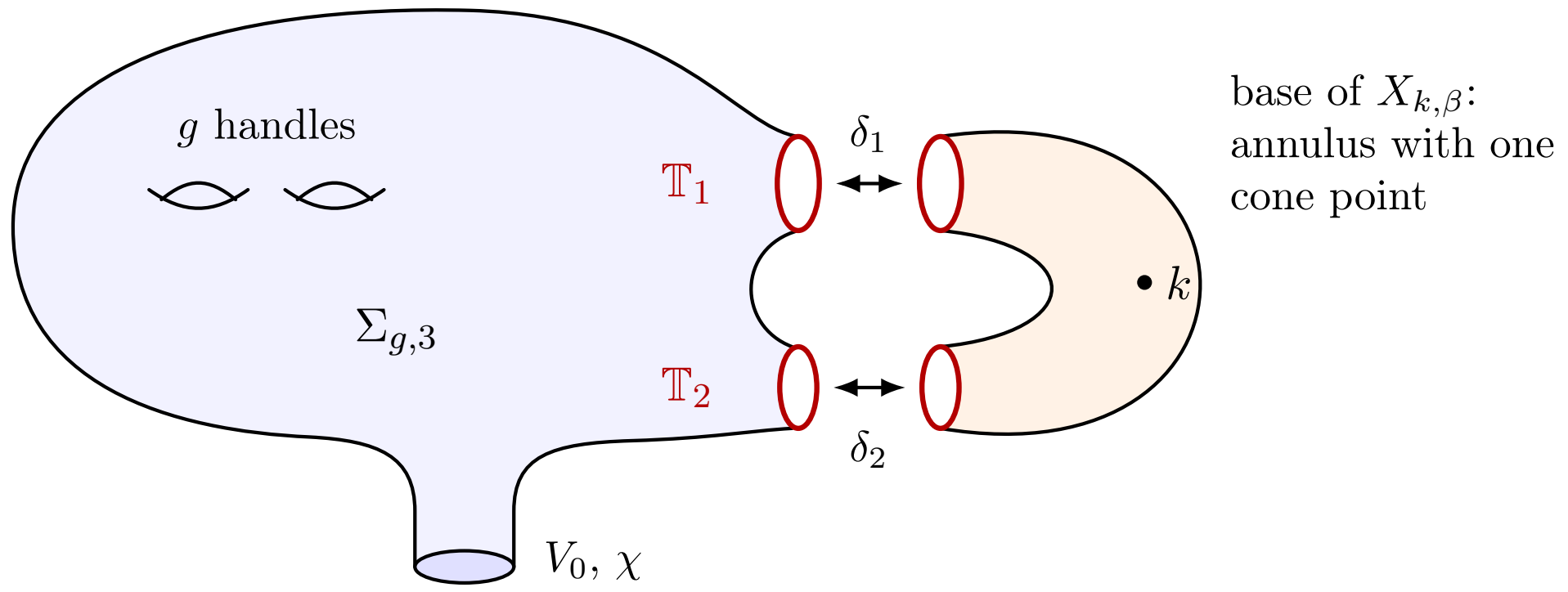
\begin{figure}[t]
  \centering
  \begin{tikzpicture}[line cap=round,line join=round,thick]
    \def\corepath{(1.65,0.98)
      .. controls (1.15,1.05) and (0.7,1.95) .. (-1.0,2.0)
      .. controls (-3.0,2.05) and (-4.7,1.55) .. (-4.7,0.25)
      .. controls (-4.7,-1.05) and (-3.4,-1.4) .. (-2.3,-1.45)
      .. controls (-1.75,-1.48) and (-1.45,-1.6) .. (-1.45,-2.05)
      -- (-1.45,-2.5) -- (-0.65,-2.5) -- (-0.65,-2.05)
      .. controls (-0.65,-1.6) and (-0.35,-1.5) .. (0.3,-1.48)
      .. controls (0.95,-1.46) and (1.25,-1.4) .. (1.6,-1.38)
      -- (1.6,-0.72)
      .. controls (1.15,-0.55) and (1.15,0.08) .. (1.65,0.22) -- cycle}
    \fill[blue!5] \corepath;
    \draw (1.65,0.98)
    .. controls (1.15,1.05) and (0.7,1.95) .. (-1.0,2.0)
    .. controls (-3.0,2.05) and (-4.7,1.55) .. (-4.7,0.25)
    .. controls (-4.7,-1.05) and (-3.4,-1.4) .. (-2.3,-1.45)
    .. controls (-1.75,-1.48) and (-1.45,-1.6) .. (-1.45,-2.05) -- (-1.45,-2.5);
    \draw (-0.65,-2.5) -- (-0.65,-2.05)
    .. controls (-0.65,-1.6) and (-0.35,-1.5) .. (0.3,-1.48)
    .. controls (0.95,-1.46) and (1.25,-1.4) .. (1.6,-1.38);
    \draw (1.6,-0.72) .. controls (1.15,-0.55) and (1.15,0.08) .. (1.65,0.22);
    \draw[very thick,red!70!black,fill=white] (1.65,0.6) ellipse (0.17 and 0.38);
    \draw[very thick,red!70!black,fill=white] (1.65,-1.05) ellipse (0.15 and 0.33);
    \node[red!70!black] at (0.75,0.6) {$\mathbb T_1$};
    \node[red!70!black] at (0.75,-1.05) {$\mathbb T_2$};
    \draw[fill=blue!12] (-1.05,-2.5) ellipse (0.4 and 0.13);
    \node[font=\small,right] at (-0.55,-2.45) {$V_0$, $\chi$};
    \foreach \x in {-3.2,-2.1}{
        \draw (\x-0.4,0.55) .. controls (\x-0.12,0.35) and (\x+0.12,0.35) .. (\x+0.4,0.55);
        \draw (\x-0.3,0.47) .. controls (\x-0.08,0.65) and (\x+0.08,0.65) .. (\x+0.3,0.47);
      }
    \node[font=\small] at (-2.65,1.05) {$g$ handles};
    \node at (-1.6,-0.6) {$\Sigma_{g,3}$};
    \path[fill=orange!10,draw] (2.8,0.98) .. controls (5.6,1.4) and (5.6,-1.85) .. (2.8,-1.38)
    -- (2.8,-0.72) .. controls (4.0,-0.6) and (4.0,0.1) .. (2.8,0.22) -- cycle;
    \draw[very thick,red!70!black,fill=white] (2.8,0.6) ellipse (0.17 and 0.38);
    \draw[very thick,red!70!black,fill=white] (2.8,-1.05) ellipse (0.15 and 0.33);
    \fill (4.45,-0.2) circle (0.06);
    \node[right] at (4.48,-0.2) {$k$};
    \node[align=left,font=\small,anchor=west] at (5.0,0.9) {base of $X_{k,\beta}$:\\annulus with one\\cone point};
    \draw[{Latex[length=2mm]}-{Latex[length=2mm]}] (1.95,0.6) -- (2.5,0.6);
    \draw[{Latex[length=2mm]}-{Latex[length=2mm]}] (1.93,-1.05) -- (2.52,-1.05);
    \node[font=\small] at (2.22,1.0) {$\delta_1$};
    \node[font=\small] at (2.22,-1.55) {$\delta_2$};
  \end{tikzpicture}
  \caption{Base orbifold of the closed manifold associated with an indecomposable Lagrangian $L_A$ for $n=2$. The boundary circles $\mathbb T_1,\mathbb T_2$ of the core base $\Sigma_{g,3}$ are joined, via $\delta_{1,2}$, by the base of the wormhole $X_{k,\beta}$, an annulus with one cone point of order $k$. When $\delta_{1,2}$ preserve the oriented fibers, the result is a Seifert manifold whose base has genus $g+1$ and one additional exceptional fiber; integral $A$ corresponds to $k=1$.}
  \label{fig:selfglue}
\end{figure}

Consequently (Figure~\ref{fig:selfglue}), the term of~\eqref{eq:common-decomp} with $L_A$ equals, up to a phase and the factor $|\Ds|^{-1/2}$, $Z_{\rm CS}$ of the closed manifold obtained from $\Mcore_{g,3}\cup_{\mathbb T_0}V_0$ by gluing $\mathbb T_1$ and $\mathbb T_2$ to the two boundaries of $X_{k,\beta}$ via $\delta_{1,2}$, times the automorphy factor. When $\delta_{1,2}$ map fibers to fibers, this manifold is Seifert fibered with one more exceptional fiber of order $k$.

For the orientable gluing $\delta_1=\delta_2=\diag(-1,1)$, for which $A=A_{\alpha,-\beta}$, the base $\Sigma_{g,3}\cup_{2\,\text{circles}}\Sigma_{0,3}$ is orientable of genus $g+1$, with two boundary circles filled by $V_0$ and the new exceptional fiber. Conversely, $A$ of the form $\bigl(\begin{smallmatrix}-1&0\\ r&1\end{smallmatrix}\bigr)$ requires the identification $\mathfrak m\mapsto-\mathfrak m'$ on one torus, so that the base is non-orientable; the corresponding partition functions are given by the core functional~\eqref{eq:core-nonor} of Section~\ref{sec:nonorientable}.

If $\delta_{1,2}$ do not map fibers to fibers, the closed manifold is obtained by gluing the Seifert pieces $\Mcore_{g,3}\cup_{\mathbb T_0}V_0$ and $X_{k,\beta}$ along the images of $\mathbb T_1$ and $\mathbb T_2$ (for $k=1$, $X_{1,\beta}\cong T^2\times I$ and a single torus suffices). It therefore carries a torus decomposition into Seifert fibered pieces, and is a graph manifold~\cite{Waldhausen1967}. For the TQFT description of gluings along torus boundaries, Seifert fibered pieces and graph manifolds, see~\cite{Angelinos:2026kxt}. Whether these tori are JSJ tori requires a further case analysis, and we do not attempt the full classification here. A torus is to be removed if the fibrations on its two sides can be matched, and parallel tori must be identified. For instance, for $g=0$ and any $\chi$, the piece $\Mcore_{0,3}\cup_{\mathbb T_0}V_0$ is $T^2\times I$, so the two gluing tori are parallel and at most one of them survives in the JSJ decomposition. In that case the closed manifold is $X_{k,\beta}$ with its two boundary tori glued to each other, or a torus bundle over the circle for $k=1$.

For indecomposable blocks of size $|B|\ge3$ we have established only the algebraic cumulant decomposition~\eqref{eq:cumulant}. One expects them to glue the corresponding boundaries of the core to each other through connected cobordisms, but the identification of these cobordisms is left open (Section~\ref{sec:discussion}).

\subsection{Two notions of connectedness}
\label{sec:twonotions}
The common-modulus average is therefore the Seifert sum plus connected contributions, and the replica average~\eqref{eq:main} isolates the sum over fillings of the core. We should distinguish two notions of connectedness here. The core is topologically connected, as reflected in the diagonal structure of $\cV$, whereas the common modulus produces statistical connectedness, as reflected in the indecomposable Lagrangians. In the model of Marolf and Maxfield~\cite{Marolf:2020xie} the two coincide, because a connected component of spacetime is a cumulant, but here they are different. The Narain modulus $m$ plays the role of the parameter $\alpha$ of~\cite{Marolf:2020xie}, and $\vev{\cdot}_{\cM_Q}$ is the analogue of their average over $\alpha$-states~\cite[(2.22)]{Marolf:2020xie}.\footnote{This analogy concerns the Narain ensemble itself, and is distinct from the formal Poisson ensemble of Section~\ref{sec:MMpoisson}, which is built from an additional sum over topologies.} The replica average gives each CFT torus its own $\alpha$-state, so the tori are connected only through the core.

Moreover, a block joining legs $i$ and $j$ glues two boundary circles of the base by an annulus. When the gluing preserves the fiber, it turns $\Sigma_{g,n}$ into $\Sigma_{g+1,n-2}$, possibly with one more cone point. A handle of the base can thus be generated by a wormhole, just as a handle of a surface in~\cite{Marolf:2020xie} can be viewed as a cylinder joining two holes, so that for a common modulus the genus of the base becomes partly dynamical.

This also bears on the question, raised in~\cite{Banerjee:2022pmw}, of whether a sum over bordisms in three dimensions can be restricted to a manageable class of 3-manifolds. As noted there for handlebodies, the requirement of gluing is a serious constraint on such a class. Our construction realizes a restricted sum of a specific kind, in which the core is fixed and only the fillings are summed, and the automorphy factors supplied by the CFT make this sum convergent for $p+q>4$. When the moduli are shared, the connected terms with $|B|=2$ glue the core to itself and, as the gluing argument above indicates, can produce graph manifolds outside the Seifert class.

\section{Seifert sums as two-dimensional topological gravity}
\label{sec:2dgrav}
In this section we reduce the construction along the fiber of the Seifert fibration. The Seifert sum then becomes a correlator in a two-dimensional TQFT, and in this form it can be compared directly with the two-dimensional topological gravity of Marolf and Maxfield~\cite{Marolf:2020xie}.

\subsection{Two-dimensional TQFT form}
\label{sec:2d}
Two-dimensional TQFTs are equivalent to commutative Frobenius algebras, i.e.\ finite-dimensional commutative unital algebras with a non-degenerate trace $\theta$~\cite{Dijkgraaf1989,Abrams:1996ty,Moore:2006dw}. Equation~\eqref{eq:core} says that reducing Abelian Chern--Simons theory along the fiber $S^1$ gives the two-dimensional TQFT of the algebra $A=\mathrm{Fun}(\Ds)$ of complex functions on $\Ds$, with pointwise multiplication. The delta functions $e_x$, defined by $e_x(y)=\delta_{x,y}$, form a basis of orthogonal idempotents, $e_xe_y=\delta_{x,y}e_x$ and $\sum_xe_x=1$, and they have trace $\theta_x:=\theta(e_x)=S_{0x}^2=|\Ds|^{-1}$. In the normalized basis $\tilde e_x:=\theta_x^{-1/2}e_x$, the genus-$g$, $n$-point amplitude is $\theta_x^{1-g-n/2}=|\Ds|^{g-1+n/2}$, which reproduces~\eqref{eq:core}. Since $2-2g-n$ is the Euler characteristic of the punctured base, this is, sector by sector, the weight $e^{S_0(2-2g-n)}$ of the topological model of~\cite{Marolf:2020xie} under the identification $e^{S_0}\leftrightarrow\theta_x^{1/2}$, and we return to this in Section~\ref{sec:MM}. The integer invariant $\chi$ inserts the twist operator $\mathsf T^{-\chi}:=\sum_x\e(-\chi q(x))e_x$. We define the \emph{Eisenstein insertion} and the \emph{theta insertion}
\begin{equation}
  \cE(\tau):=\sum_{x\in\Ds}E_x(\tau)\,\tilde e_x,\qquad
  \vartheta(\tau;m):=\sum_{x\in\Ds}\Theta_x(\tau;m)\,\tilde e_x\;.
\end{equation}
Then~\eqref{eq:main} becomes
\begin{equation}
  \cZ_{g,n;\chi}(\btau)=\bigl\langle\cE(\tau_1)\cdots\cE(\tau_n)\,\mathsf T^{-\chi}\bigr\rangle_{\Sigma_g}
    =\Bigl\langle\bigl\langle\vartheta(\tau_1;m_1)\cdots\vartheta(\tau_n;m_n)\,\mathsf T^{-\chi}\bigr\rangle_{\Sigma_g}\Bigr\rangle_{(\cM_Q)^n}\;.
\end{equation}
Each exceptional fiber is a local operator on the base orbifold $\Sigma_g$, and the ensemble average turns the theta insertion into the Eisenstein insertion, which is the generating function of all fillings of that fiber.

\subsection{Dictionary with the Marolf--Maxfield model}
\label{sec:MM}
We now compare the Seifert sum with the two-dimensional topological gravity of Marolf and Maxfield~\cite{Marolf:2020xie}. There, spacetimes are oriented surfaces weighted by $e^{S_0}$ to the power of their Euler characteristic, and each boundary circle carries an observable $Z$. It is interpreted as $Z=\operatorname{Tr}\mathbbm 1$, the dimension of the Hilbert space of a dual topological quantum mechanics. After reduction along the fiber, the base of the core is such a surface, and the interface tori project to its boundary circles. Table~\ref{tab:MM} summarizes the dictionary, which we now explain.

\paragraph{Weights and boundary observables.}
By Section~\ref{sec:2d}, in each sector $x$ the core carries the weight of~\cite{Marolf:2020xie} under the identification
\begin{equation}
  e^{S_0}\ \longleftrightarrow\ \theta_x^{1/2}=S_{00}=|\Ds|^{-1/2}\;.
  \label{eq:MM-S0}
\end{equation}
Marolf and Maxfield also introduce a boundary parameter $S_\partial$. They set $S_\partial=S_0$, which removes the boundaries from the Euler characteristic and amounts to rescaling $Z$ by $e^{S_\partial}$. In our case the same rescaling is implemented by the normalized idempotents $\tilde e_x$. In their convention the boundary observable of sector $x$ is
\begin{equation}
  Z\ \longleftrightarrow\ \mathsf Z_x(\tau):=|\Ds|^{1/2}E_x(\tau)\;,
  \label{eq:MM-Z}
\end{equation}
or $|\Ds|^{1/2}\Theta_x(\tau;m)$ before the average. There are three differences. First, there are $|\Ds|$ sectors instead of one. The idempotents $e_x$ play the role of the disjoint ``universes'' of~\cite{Banerjee:2022pmw}, in that a connected bordism is non-zero only if all its insertions lie in a single sector, which is the diagonal structure of~\eqref{eq:core}. Second, $S_0$ is fixed by the TQFT instead of being a free parameter. Third, the boundary circle times the fiber is a torus carrying the Narain theory, so that the topological quantum mechanics of~\cite{Marolf:2020xie} is replaced by a two-dimensional CFT, with $\Theta_x$ its partition function in the sector where the anyon $x$ threads the fiber.

After the average, $E_x$ is the sum over all ways of capping the boundary circle. A solid torus $V_{(c,d)}$ projects to a disc with a cone point of order $c$, which is smooth for $c=1$, while for $c=0$ the fiber collapses.

\paragraph{Generating functions.}
The connected weight factorizes into a genus part $\theta_x^{1-g}$ and a factor for each boundary. This explains the exponentiation in~\cite[(3.14)]{Marolf:2020xie}, and the generating functions at fixed genus are parallel:
\begin{equation}
  \sum_{n}\frac{u^n}{n!}\,e^{S_0(2-2g)}Z^n=e^{S_0(2-2g)}\,e^{uZ}
  \qquad\longleftrightarrow\qquad
  \mathcal F_{g,\chi}(u;\tau)=\sum_{x\in\Ds}\theta_x^{1-g}\,\e(-\chi q(x))\,e^{u\,\mathsf Z_x(\tau)}\;.
\end{equation}
In~\cite{Banerjee:2022pmw} the model of~\cite{Marolf:2020xie} is recovered from the Hartle--Hawking covector evaluated on the exponentiated diagonal. Here the exponent is replaced by the Eisenstein insertion $u\,\mathsf Z_x(\tau)$, so the ensemble average supplies the boundary data. The labeled and unlabeled counting of Section~\ref{sec:egf} parallels their distinguishable boundaries and symmetry factors.

\begin{table}[t]
  \centering
  \setlength{\abovecaptionskip}{0pt}
  \setlength{\belowcaptionskip}{8pt}
  \caption{Dictionary between the topological model of~\cite{Marolf:2020xie} and the Seifert sum. Entries marked ``Narain'' refer to the common-modulus Narain ensemble of Section~\ref{sec:common}, and the entry marked ``Poisson'' to the formal ensemble built from the additional topology sum in Section~\ref{sec:MMpoisson}. The two ensembles are different and are not identified with each other.}
  \label{tab:MM}
  \small
  \begin{tabular}{@{}>{\raggedright\arraybackslash}p{3.9cm}>{\raggedright\arraybackslash}p{4.6cm}>{\raggedright\arraybackslash}p{6.2cm}@{}}
    \toprule
                                          & Marolf--Maxfield~\cite{Marolf:2020xie}   & this paper (reduced along the fiber)                                                              \\
    \midrule
    spacetime                             & oriented surface                             & base $\Sigma_{g,n}$ of the core, with caps                                                        \\
    two-dimensional TQFT                               & trivial, $e^{S_0\times(\text{Euler char.})}$ & $\mathrm{Fun}(\Ds)$, $|\Ds|$ sectors, $e^{S_0}\leftrightarrow|\Ds|^{-1/2}$                                      \\
    boundary                              & circle, $Z=\dim\cH$                          & torus (circle $\times$ fiber), $|\Ds|^{1/2}\Theta_x(\tau;m)$                                      \\
    averaged boundary                     & ---                                          & $\mathsf Z_x=|\Ds|^{1/2}E_x(\tau)$: sum over caps                                                         \\
    $\log$ of unnormalized gen.\ function & $\lambda e^{u}$                              & $\lambda_{\rm eff}\sum_xe^{u\mathsf Z_x}$, \eqref{eq:logG}                                                 \\
    genus-summed coupling                 & $\lambda=1/(y(1-y))$, $y=e^{-2S_0}$          & $\lambda_{\rm eff}$, $y_{\rm eff}=|\Ds|e^{-2\hat S_0}$, \eqref{eq:lambdaeff}                           \\
    moments                               & partitions of boundaries                     & partitions of legs, \eqref{eq:cumulant} (Narain, Section~\ref{sec:common})                           \\
    disc / cylinder                       & disc / cylinder                              & solid tori $V_{(c,d)}$ / wormholes $T^2\times I$, $X_{k,\beta}$                                   \\
    handle                                & part of the genus sum                        & wormhole between two legs (Section~\ref{sec:selfglue})                                            \\
    $\alpha$-parameters                   & $\alpha$                                     & Narain modulus $m$ (Narain, Section~\ref{sec:common})                                                \\
    baby-universe $\alpha$-states         & $\ket{Z=d}$, $d\in\NN$                       & $\ket{\{N_x\}}$, $N_x\in\NN$ (Poisson, Section~\ref{sec:MMpoisson}; formal, not identified with $m$) \\
    unimodular case                       & ---                                          & same formal generating function with genus weight, $Z=d\to d\cdot E(\tau)$                                                       \\
    \bottomrule
  \end{tabular}
\end{table}

\subsection{Sum over the genus}
\label{sec:MMgenus}
The two settings differ in the sum over the genus of the base. All $\theta_x=|\Ds|^{-1}$ are at most one, so $\sum_g|\Ds|^{g-1}$ diverges. For $n=0$ this is the divergence of the sub-sum $\sum_g\dim\cH(\Sigma_g)$ over $\Sigma_g\times S^1$, which~\cite{Banerjee:2022pmw} identify as a basic obstruction to summing over bordisms in three dimensions. Let us now weight each connected base by $e^{\hat S_0(2-2g)}$, as in~\cite{Marolf:2020xie}, with $\hat S_0$ a free parameter, not to be confused with the fixed value~\eqref{eq:MM-S0}. The genus sum then gives
\begin{equation}
  \lambda_{\rm eff}:=\sum_{g\ge0}e^{\hat S_0(2-2g)}|\Ds|^{g-1}=\frac{1}{y_{\rm eff}(1-y_{\rm eff})},\qquad y_{\rm eff}:=|\Ds|\,e^{-2\hat S_0}\;.
  \label{eq:lambdaeff}
\end{equation}
This is the connected amplitude $\lambda=1/(y(1-y))$, $y=e^{-2S_0}$, of~\cite[(3.9)]{Marolf:2020xie}, with $y\to y_{\rm eff}$. The discriminant group thus renormalizes $e^{2\hat S_0}\to e^{2\hat S_0}/|\Ds|$. The sum converges for $|y_{\rm eff}|<1$, and $\lambda_{\rm eff}$ is invariant under $y_{\rm eff}\to1-y_{\rm eff}$, which is the analogue of the strong--weak duality of~\cite[(3.20)]{Marolf:2020xie}. Pure Chern--Simons theory corresponds to $e^{\hat S_0}=1$, i.e.\ $y_{\rm eff}=|\Ds|\ge1$, which lies outside the region of convergence.

\subsection{A formal Poisson ensemble}
\label{sec:MMpoisson}
This subsection is formal and is not used elsewhere in the paper. It transcribes the baby-universe construction of~\cite[Section~3]{Marolf:2020xie} to the weighted genus sum of Section~\ref{sec:MMgenus}, and its caveats are collected at the end.

We now include disconnected bases with the symmetry factors of~\cite{Marolf:2020xie}. The Poisson ensemble obtained in this way is a second, formal ensemble, built from the averaged amplitudes by the additional genus weight and the exponentiation over disconnected components. We do not identify the Narain measure on $\cM_Q$ with a Poisson distribution, nor do we claim any map $m\mapsto\{N_x\}$ relating the two. For instance, extending the generating function below to sources $u_i$ at different $\tau_i$, the connected two-point function of the Poisson ensemble is $\lambda_{\rm eff}\sum_x\mathsf Z_x(\tau_1)\mathsf Z_x(\tau_2)$. In the common-modulus Narain ensemble of Section~\ref{sec:common}, the corresponding cumulant is instead $\vev{\Theta_x(\tau_1;m)\Theta_y(\tau_2;m)}_{\rm c}=E^{[2],\rm conn}_{x,y}(\tau_1,\tau_2)$, which is a different function.

In the following we restrict to the subfamily with $\chi=0$.\footnote{The regular fibers in $E_x$ produce manifolds with other values of $\chi$, but the weighted generating functions at different fixed $\chi$ are not the same. For $n=0$, for instance, $\chi$ enters through $\sum_x\e(-\chi q(x))$.} Each connected component chooses its own sector, so, as in~\cite[(3.12)--(3.14)]{Marolf:2020xie}, the generating function of $n$ fibers at a common $\tau$ satisfies
\begin{equation}
  \log\mathcal G(u;\tau)=\lambda_{\rm eff}\sum_{x\in\Ds}e^{u\,\mathsf Z_x(\tau)}\;,
  \label{eq:logG}
\end{equation}
where $\mathcal G(u;\tau)$ denotes the unnormalized generating function, summed over possibly disconnected bases. Its value $\mathcal G(0;\tau)=e^{|\Ds|\lambda_{\rm eff}}$ is the contribution of closed manifolds without fibers. After division by it, we obtain the normalized expectation value
\begin{equation}
  \log\mathbb E\bigl[e^{u\widehat Z(\tau)}\bigr]=\log\frac{\mathcal G(u;\tau)}{\mathcal G(0;\tau)}=\lambda_{\rm eff}\sum_{x\in\Ds}\bigl(e^{u\,\mathsf Z_x(\tau)}-1\bigr)\;,
\end{equation}
which is the moment generating function of
\begin{equation}
  \widehat Z(\tau):=\sum_{x\in\Ds}N_x\,\mathsf Z_x(\tau),\qquad N_x\ \text{independent Poisson variables with mean }\lambda_{\rm eff}\;.
  \label{eq:MM-poisson}
\end{equation}
Formally, the baby-universe $\alpha$-states $\ket{Z=d}$ of~\cite{Marolf:2020xie} become $\ket{\{N_x\}_{x\in\Ds}}$, where $N_x$ counts the connected components whose fiber holonomy is $x$. The Hartle--Hawking state is a product of $|\Ds|$ coherent states.

The labels $\{N_x\}$ are, however, an extended formal labeling, and they are not all distinguished by the boundary observables $\widehat Z(\tau)$. Indeed, by~\eqref{eq:E-aut}, $\mathsf Z_{ax}=\mathsf Z_x$ for every $a\in\Aut(\Ds,q)$; in particular $\mathsf Z_{-x}=\mathsf Z_x$. Hence $\widehat Z(\tau)$, for all $\tau$, depends only on the orbit sums $M_{\mathcal O}=\sum_{x\in\mathcal O}N_x\sim\mathrm{Poisson}(|\mathcal O|\lambda_{\rm eff})$ over the $\Aut(\Ds,q)$-orbits $\mathcal O$. For $\Ds=\ZZ_3$, for example, $\widehat Z=N_0\mathsf Z_0+(N_1+N_2)\mathsf Z_1$, so $(N_1,N_2)=(1,0)$ and $(0,1)$ cannot be distinguished. Resolving the individual $N_x$ would require additional sector-resolved observables. This is a degeneracy of the observables considered here, and should not be confused with the null states of~\cite{Marolf:2020xie}. However, $\mathsf Z_x(\tau)$ is complex in general, and we have not examined reflection positivity, which plays a central role in~\cite{Marolf:2020xie} in fixing $S_\partial$ and in deriving the null states. The statements of this subsection should therefore be regarded as formal.

\paragraph{Unimodular lattices.}
For even unimodular $\Lambda$, for instance $\mathrm{II}_{c,c}$ with $c>2$ (the original Narain ensemble), $\Ds$ is trivial and $\cV_{g,n;\chi}=1$. Then $\cZ_{g,n;\chi}(\btau)=\prod_iE(\tau_i)$, and~\eqref{eq:logG}--\eqref{eq:MM-poisson} become
\begin{equation}
  \log\mathcal G(u;\tau)=\lambda\,e^{u\,E(\tau)},\qquad
  \log\mathbb E\bigl[e^{u\widehat Z(\tau)}\bigr]=\lambda\bigl(e^{u\,E(\tau)}-1\bigr),\qquad \widehat Z(\tau)=d\cdot E(\tau)\;,
\end{equation}
with $d$ Poisson distributed with the \emph{same} mean $\lambda$ as in~\cite{Marolf:2020xie}, where $\lambda=\lambda_{\rm eff}$ for $|\Ds|=1$. With the additional genus weight $e^{\hat S_0(2-2g)}$, this formal generating function is that of their model with $Z=d$ replaced by $d\cdot E(\tau)$. Their integer $d$ is the value of $Z$, interpreted there as the dimension of a dual Hilbert space. Here it multiplies $E(\tau)$, the sum over fillings contributed by each connected Seifert component. For $|\Ds|>1$ the anyons produce the sectors $x$, the twists $\e(-\chi q(x))$, and the renormalization of $\hat S_0$.

\subsection{What the parallel does not capture}
\label{sec:bulk}
The parallel with~\cite{Marolf:2020xie} is not exact. To begin with, the argument goes in the opposite direction. Marolf and Maxfield derive an ensemble from a sum over topologies, whereas we start from the Narain ensemble and derive a sum over topologies. The two points of view meet in the statement that the Narain modulus $m$ plays the role of the parameter $\alpha$ (Section~\ref{sec:twonotions}). The Poisson ensemble of Section~\ref{sec:MMpoisson} is a separate, formal construction, and we have not related it to the Narain measure.

Also, the CFT tori are interfaces and not asymptotic boundaries. In the ``bulk'' of~\cite{Ashwinkumar:2021kav}, the CFT torus is the conformal boundary of the solid tori $V_{(c,d)}$, and the second solid torus which closes the lens space is a formal device encoding the CFT label $h$. Here it is replaced by an actual piece of the geometry, the core, which is shared by the $n$ CFT tori. The closed Seifert manifold should therefore not be thought of as the Euclidean spacetime of a gravitational path integral: it is obtained by gluing the summed pieces $V_i$ to the topological sector that labels the CFT observable. For this reason the comparison with~\cite{Marolf:2020xie} is most direct at the level of the base. The issue of the gauge group, $U(1)^{p+q}$ versus $\RR^{p+q}$~\cite[Section~2.4]{Ashwinkumar:2021kav}, applies here as well, and we have used $U(1)^{p+q}$, so that the boundary Hilbert space is $\CC[\Ds]$.

Several features of the construction are also specific to three dimensions. The $\SL(2,\ZZ)$ family of caps, the degenerate fillings, and the graph manifolds produced by self-gluings which do not preserve the fiber (Section~\ref{sec:selfglue}) have no two-dimensional counterpart. Exceptional fibers resemble the conical defects and orbifold loci of~\cite{Benjamin:2020mfz,Maxfield:2020ale}. The $\ZZ_m$ orbifolds of~\cite{Maloney:2007ud}, which are Seifert fibered orbifolds rather than smooth Seifert manifolds, are discussed in Appendix~\ref{sec:orbifolds}.

Finally, we have no counterpart yet of the end-of-the-world branes of~\cite{Marolf:2020xie}, nor of the resulting bound on the rank of their inner products and the associated null states.

\section{Discussion and open problems}
\label{sec:discussion}

Our construction is based on the fact that Abelian Chern--Simons theory on $\Sigma_{g,n}\times S^1$ is diagonal in the fiber holonomy, so that a Seifert manifold is a genus-$g$ correlator of local fiber insertions in a finite two-dimensional TQFT. The Narain ensemble average replaces each theta insertion by an Eisenstein insertion, the generating function of all fillings of that fiber, and modularity follows from the Weil representation on each leg: the modular $T$-transformation is the Seifert move, and the $S$-transformation relabels slopes.

The weights~\eqref{eq:ZM} of the resulting Seifert sum agree with the surgery formula, including orientation and framing phase~\eqref{eq:ZMcan}. For a common modulus the average acquires connected terms labeled by indecomposable Lagrangians, and for two boundaries each of them is realized by a wormhole, whose topology is not fixed uniquely by the Lagrangian. After reduction along the fiber, the construction closely resembles the model of Marolf and Maxfield~\cite{Marolf:2020xie}, with $e^{S_0}\leftrightarrow|\Ds|^{-1/2}$, and for unimodular lattices, once a genus weight is introduced, the formal generating functions of the two coincide. Since modular invariance forces the sum to include the degenerate fillings, which need not be Seifert manifolds, modularity could be an organizing principle in the summation of geometries in quantum gravity, in that it tells us which geometries must be included.

The most immediate direction for future work is to carry out the sum over topologies itself. One would choose weights for $g$, $n$ and $\chi$, for instance in~\eqref{eq:egf}, and study the convergence of the full sum over $g$ and $\chi$. The divergence in $g$ is the one discussed in~\cite{Banerjee:2022pmw} (see Section~\ref{sec:MMgenus}), and it would be interesting to see whether their splitting formula has an analogue for the Seifert sums. Since the Abelian partition functions depend on the closed manifold only through its first homology and linking form, cf.~\eqref{eq:linking}, the convergent measures on homology groups constructed in~\cite{Nicosanti:2025xwu} may provide a natural way of organizing such a sum.

The $\alpha$-states $\ket{\{N_x\}}$ of Section~\ref{sec:MMpoisson} are only formal, and it would be important to make them precise. This requires checking reflection positivity, and asking whether the rank bound and the null states of~\cite{Marolf:2020xie} have Seifert counterparts. One would also like to find the analogue of the end-of-the-world branes of~\cite{Marolf:2020xie}, which might be bases with boundary circles on which the TQFT ends on a topological boundary condition $\Psi_H$ (Section~\ref{sec:dehn}).

We have interpreted the connected terms geometrically only for two boundaries. It remains to extend the self-gluing interpretation of Section~\ref{sec:selfglue} to indecomposable blocks with $|B|\ge3$, and to identify the corresponding multi-boundary cobordisms.

On the modular side, the constant-term computation of Section~\ref{sec:example} could be extended to the full Fourier expansions of $\cZ_{g,n;\chi}$ and of the kernels $K_{\mathfrak s;\chi}$, for general $\Ds$, and one may ask whether they admit a spectral interpretation analogous to~\cite{Keller:2014xba}. The Hecke-type observable of Appendix~\ref{sec:orbifolds} should likewise be made precise, together with its Poincar\'e expansion over the orbifolds $M_{(c,d)}/\ZZ_m$.

One can also consider a larger class of theories. For odd $p+q$, $\rho$ is a representation of $\Mp(2,\ZZ)$, and~\eqref{eq:ZMcan} has to be extended with the corresponding eighth-root phases. For odd $Q$ one needs spin Chern--Simons theory~\cite[Section~3]{Ashwinkumar:2021kav}, spin structures on Seifert manifolds, and a fiber basis compatible with them. Going beyond the Abelian case, one can consider Poincar\'e sums for rational CFTs~\cite{Castro:2011zq,Meruliya:2021utr,Raeymaekers:2021ypf}. There the analogue of~\eqref{eq:core} is the Lawrence--Rozansky formula~\cite{LawrenceRozansky1999}, and the two-dimensional TQFT is that of the Verlinde algebra, viewed as a commutative Frobenius algebra. Its idempotents $e_j$ have traces $\theta_j=S_{0j}^2$, which in general depend on $j$, unlike the constant traces $\theta_x=|\Ds|^{-1}$ of the Abelian case.

Finally, one would like to understand the structural origin of the core functional. The discussion in~\cite[Section~5]{Ashwinkumar:2021kav}, and the recent derivation of code-CFT ensemble holography from Howe duality over finite fields~\cite{Dymarsky:2025agh}, suggest viewing $\cV_{g,n;\chi}$ as selecting a vector in the $O(p,q)\times\Sp(2n)$ Howe-dual picture. It would be interesting to see whether the diagonal structure~\eqref{eq:core} is natural from this point of view.

\section*{Acknowledgments}
\addcontentsline{toc}{section}{Acknowledgments}
This research was supported in part by the World Premier International Research Center Initiative (WPI), MEXT, Japan; JSPS KAKENHI Grant No.~23K25865; JST, Japan (CREST Grant No.~JPMJCR26XA, Moonshot R\&D Grant No.~JPMJMS256E); and IBM-UTokyo-sponsored research. The authors were assisted by Anthropic Claude and OpenAI ChatGPT in writing and editing the manuscript.

\appendix
\section{Surgery formula, framing phase, and the Dedekind eta factors}
\label{app:surgery}

\subsection{Derivation of the framed formula~\texorpdfstring{\eqref{eq:ZMcan}}{}}
\label{sec:ZMcanderiv}

\paragraph{Surgery presentation.}
Recall that $(r/s)$-surgery on a knot $K\subset S^3$ removes a tubular neighborhood $N(K)$ and glues back a solid torus whose meridian is attached along $r\mathfrak m_K+s\mathfrak l_K$, where $\mathfrak m_K$ is the meridian of $K$ and $\mathfrak l_K$ is its Seifert longitude. The surgery is integral if $s=1$, and surgery on a link is performed on each component.

\begin{figure}[t]
  \centering
  \begin{tikzpicture}[line cap=round,thick,baseline=(current bounding box.center)]

    \draw (0,0) ellipse (3.4 and 1.3);
    \node at (-3.75,0) {$\chi$};
    \node at (-3.35,1.05) {$K_0$};
    \foreach \t/\lab in {230/{-\alpha_1/\beta_1},262/{-\alpha_2/\beta_2},302/{-\alpha_n/\beta_n}}{
        \pgfmathsetmacro{\cx}{3.4*cos(\t)}
        \pgfmathsetmacro{\cy}{1.3*sin(\t)}
        \draw (\cx,\cy) ellipse (0.26 and 0.55);
        \draw[white,line width=5pt] plot[domain=\t-7:\t-1,samples=12] ({3.4*cos(\x)},{1.3*sin(\x)});
        \draw plot[domain=\t-9:\t+1,samples=16] ({3.4*cos(\x)},{1.3*sin(\x)});
        \node at (\cx,-2.25) {$\lab$};
      }
    \node at (0.65,-2.25) {$\cdots$};
    \foreach \t/\ang/\lab in {115/10.1/{B_1},65/-10.1/{B_g}}{
        \node[draw,fill=white,rotate=\ang,minimum width=0.9cm,minimum height=0.55cm,inner sep=1pt] at ({3.4*cos(\t)},{1.3*sin(\t)}) {$\lab$};
      }
    \foreach \t in {78,90,102}{
        \fill ({3.4*cos(\t)},{1.3*sin(\t)+0.17}) circle (0.03);
      }
    \node at (0,-2.85) {(a)};
  \end{tikzpicture}
  \qquad
  \begin{tikzpicture}[line cap=round,thick,scale=0.85,baseline=(current bounding box.center)]
    \draw[dashed,gray] (-2.05,-1.75) rectangle (2.05,1.75);
    \draw[red!70!black,very thick] (-2.05,-1.5) to[out=0,in=295] ([shift={(0,-0.620)}]205:1) arc (205:-25:1) to[out=-115,in=180] (2.05,-1.5);
    \draw (-0.537,0.310) circle (1);
    \draw (0.537,0.310) circle (1);
    \draw[white,line cap=butt,line width=4.2pt] ([shift={(-0.537,0.310)}]46.5:1) arc (46.5:68.5:1);
    \draw ([shift={(-0.537,0.310)}]43.5:1) arc (43.5:71.5:1);
    \draw[white,line cap=butt,line width=4.2pt] ([shift={(-0.537,0.310)}]291.5:1) arc (291.5:313.5:1);
    \draw ([shift={(-0.537,0.310)}]288.5:1) arc (288.5:316.5:1);
    \draw[white,line cap=butt,line width=4.2pt] ([shift={(0.537,0.310)}]286.5:1) arc (286.5:308.5:1);
    \draw ([shift={(0.537,0.310)}]283.5:1) arc (283.5:311.5:1);
    \draw[white,line cap=butt,line width=4.2pt] ([shift={(0.537,0.310)}]171.5:1) arc (171.5:193.5:1);
    \draw ([shift={(0.537,0.310)}]168.5:1) arc (168.5:196.5:1);
    \draw[white,line cap=butt,line width=4.2pt] ([shift={(0,-0.620)}]166.5:1) arc (166.5:188.5:1);
    \draw[red!70!black,very thick] ([shift={(0,-0.620)}]163.5:1) arc (163.5:191.5:1);
    \draw[white,line cap=butt,line width=4.2pt] ([shift={(0,-0.620)}]51.5:1) arc (51.5:73.5:1);
    \draw[red!70!black,very thick] ([shift={(0,-0.620)}]48.5:1) arc (48.5:76.5:1);
    \node at (-1.72,1.42) {$a_j$};
    \node at (1.72,1.42) {$b_j$};
    \node[red!70!black] at (-1.65,-1.2) {$K_0$};
    \node at (0,-2.2) {(b)};
  \end{tikzpicture}
  \caption{(a) Surgery presentation of $M(g;\chi;(\alpha_i,\beta_i))$: an unknot $K_0$ with coefficient $\chi$, $n$ meridians of $K_0$ with rational coefficients $-\alpha_i/\beta_i$, and $g$ boxes $B_j$. (b) The content of one box $B_j$: the arc of $K_0$ in the box and a pair of $0$-framed unknots $a_j,b_j$, which form Borromean rings with $K_0$. The pairs $(a_j,b_j)$ contribute only zero rows and columns to the linking matrix. The orientation-reversed manifold $-M$ is obtained by reversing all coefficients.}
  \label{fig:surgery}
\end{figure}
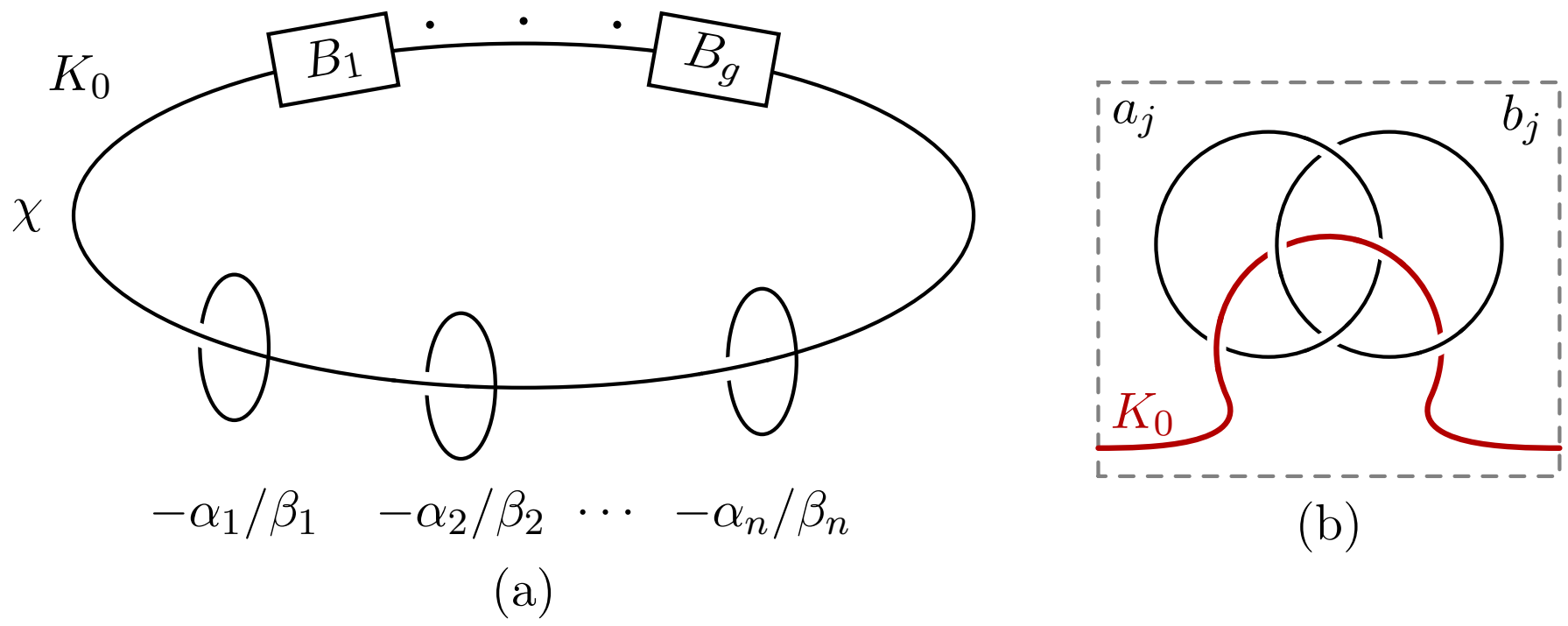

Let $K_0\subset S^3$ be an unknot and $K_1,\dots,K_n$ meridian circles of $K_0$, oriented so that $\mathrm{lk}(K_i,K_0)=1$. The complement of $K_0$ is an open solid torus whose core is parallel to the $K_i$. Removing the $K_i$ leaves $\Sigma_{0,n+1}\times S^1$, where the fiber is parallel to the $K_i$ and the base is oriented so that base $\times$ fiber has the orientation of $S^3$. On the outer boundary torus $\mathbb T_0=\partial N(K_0)$, the section boundary is the Seifert longitude and the fiber is the meridian: $\mathfrak l_0=\mathfrak l_{K_0}$ and $\mathfrak m_0=\mathfrak m_{K_0}$. On the inner tori, the section boundary has the clockwise induced orientation: $\mathfrak l_i=-\mathfrak m_{K_i}$ and $\mathfrak m_i=\mathfrak l_{K_i}$, where the parallel copies of the core are unlinked from each other. The two roles are exchanged because the fiber bounds a disc in $N(K_0)$, whereas the section boundary bounds a disc in $N(K_i)$. Gluing $V_0$ along $\mathfrak l_0+\chi \mathfrak m_0=\mathfrak l_{K_0}+\chi\mathfrak m_{K_0}$ is $\chi$-surgery on $K_0$. Gluing $V_i$ along $\alpha_i\mathfrak l_i+\beta_i\mathfrak m_i=-\alpha_i\mathfrak m_{K_i}+\beta_i\mathfrak l_{K_i}$ is $(-\alpha_i/\beta_i)$-surgery on $K_i$. For $g>0$, the standard presentation~\cite[Figures~5.25 and~6.4]{GompfStipsicz1999} adds $g$ pairs $(a_j,b_j)$ of $0$-framed unknots, one pair in each box $B_j$ along $K_0$ (Figure~\ref{fig:surgery}). Each triple $(a_j,b_j,K_0)$ forms the Borromean rings: any two of the three components are unlinked, but the three together cannot be separated. Performing $0$-surgery on $a_j$ and $b_j$ adds one handle to the base. To see this, recall that $0$-surgery on all three Borromean rings gives $T^3$, in which the meridian of one component is a fiber $S^1$. Thus $0$-surgery on two components alone turns the complement of the third into $\Sigma_{1,1}\times S^1$, whose fiber is the meridian of $K_0$, in agreement with the case $g=0$ above. The $a_j$ and $b_j$ have framing $0$ and vanishing linking numbers with all the other components, so they contribute only zero rows and columns to the linking matrix. In conclusion, $M(g;\chi;(\alpha_i,\beta_i))$ is the surgery $(\chi;-\alpha_i/\beta_i)$, and $-M$ is the surgery $(-\chi;\alpha_i/\beta_i)$.

\paragraph{Gauss sum.}
We evaluate~\eqref{eq:surgery} for $-M$, assuming $\alpha_i\ge1$. We expand $\alpha_i/\beta_i=a_{i1}-1/(a_{i2}-1/(\cdots-1/a_{i\ell_i}))$ with $a_{ij}\ge2$ for $j\ge2$. For $\beta_i=0$, so that $\alpha_i=1$, the surgery on $K_i$ is trivial, and we take the chain to be empty, with $\ell_i=0$, $\det\mathsf C_i=1$, $(\mathsf C_i^{-1})_{11}=0$ and $\sigma(\mathsf C_i)=0$; the formulas below then hold as they stand. The linking matrix $W$ is then star-shaped: the center has entry $-\chi$, it is joined to the first entry of each of the $n$ chains $\mathsf C_i$ (tridiagonal, with diagonal entries $a_{ij}$ and off-diagonal entries $1$), and there are $2g$ zero rows,
\[
  W=\begin{pmatrix}
    -\chi  & v^T    & \cdots & v^T    & 0      \\
    v      & \mathsf C_1    &        &        &        \\
    \vdots &        & \ddots &        &        \\
    v      &        &        & \mathsf C_n    &        \\
    0      &        &        &        & 0_{2g}
  \end{pmatrix},\qquad
  \vcenter
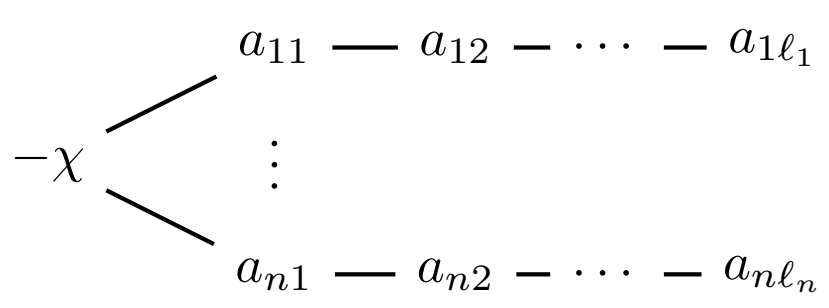
{\hbox{\begin{tikzpicture}[thick,every node/.style={font=\small}]
        \node (c) at (0,0) {$-\chi$};
        \node (a1) at (1.5,0.75) {$a_{11}$};
        \node (b1) at (2.7,0.75) {$a_{12}$};
        \node (d1) at (3.7,0.75) {$\cdots$};
        \node (e1) at (4.8,0.75) {$a_{1\ell_1}$};
        \node at (1.5,0.08) {$\vdots$};
        \node (an) at (1.5,-0.75) {$a_{n1}$};
        \node (bn) at (2.7,-0.75) {$a_{n2}$};
        \node (dn) at (3.7,-0.75) {$\cdots$};
        \node (en) at (4.8,-0.75) {$a_{n\ell_n}$};
        \draw (c) -- (a1) -- (b1) -- (d1) -- (e1);
        \draw (c) -- (an) -- (bn) -- (dn) -- (en);
      \end{tikzpicture}}}
\]
where $v:=(1,0,\dots,0)^T$. In the graph, the vertices are the non-zero rows, labeled by their diagonal entries, and the edges are the off-diagonal entries $1$. Each zero row contributes a factor $|\Ds|$. The matrix $B_{xy}:=\e((x,y))$ and the vector $\sum_y\ket y$ satisfy
\begin{equation}
  B=c\,\rho(S)^{-1},\qquad \sum_y\ket y=c\,\rho(S)^{-1}\ket0,\qquad c:=|\Ds|^{1/2}\e(-\sigma/8)\;.
\end{equation}
Hence summing over the variables of the $i$-th chain, with the center variable $x$ fixed, gives
\begin{equation}
  c^{\,\ell_i+1}\,\bra x\rho(w_i)\ket0,\qquad w_i:=S^{-1}T^{a_{i1}}S^{-1}T^{a_{i2}}\cdots S^{-1}T^{a_{i\ell_i}}S^{-1}\;.
\end{equation}
The matrix $T^{a_{i1}}S\cdots T^{a_{i\ell_i}}S$ has first column $\epsilon_i(\alpha_i,\beta_i)$, where $\epsilon_i:=\operatorname{sign}(\det \mathsf C_i)$ (its first entry is the continuant $\det \mathsf C_i$). Since $S^{-1}=-S$, $w_i$ has first column $s_i(\beta_i,-\alpha_i)$ with $s_i:=(-1)^{\ell_i}\epsilon_i$. Thus $w_i=s_i\gamma_i^{-1}T^k$, where $\gamma_i$ has lower row $(\alpha_i,\beta_i)$. Using $\rho(T)\ket0=\ket0$ and $\rho(-\mathbbm 1)\ket0=\rho(S)^2\ket0=\e(-\sigma/4)\ket0$, we get $\bra x\rho(w_i)\ket0=\e(-\sigma t_i/4)\,\varepsilon_{Q,x}(\alpha_i,\beta_i)$ with $t_i:=(1-s_i)/2$. The powers of $|\Ds|$ combine to $|\Ds|^{g-1+n/2}$. Therefore
\begin{equation}
  Z^{\rm can}_{\rm CS}[-M]=\e\Bigl(-\frac{\sigma\kappa}{8}\Bigr)|\Ds|^{g-1+\frac n2}\sum_x\e(-\chi q(x))\prod_i\varepsilon_{Q,x}(\alpha_i,\beta_i),\qquad
  \kappa:=\sigma(W)+\sum_{i=1}^n(\ell_i+1+2t_i)\;.
\end{equation}

\paragraph{The phase.}
Since $(\mathsf C_i^{-1})_{11}=\beta_i/\alpha_i$, Sylvester's law of inertia applied to the Schur complement of the chains gives $\sigma(W)=\sum_i\sigma(\mathsf C_i)+\operatorname{sign}(-\chi-\sum_i\beta_i/\alpha_i)=\sum_i\sigma(\mathsf C_i)-\operatorname{sign}(\mathsf e)$. This also holds for $\mathsf e=0$. Let $p_i$ be the number of positive eigenvalues of $\mathsf C_i$, which is non-degenerate since $|\det \mathsf C_i|=\alpha_i$. Then $\sigma(\mathsf C_i)+\ell_i=2p_i$, and $(-1)^{\ell_i-p_i}=\epsilon_i$, so $(-1)^{p_i}=s_i=(-1)^{t_i}$. Hence $\sigma(\mathsf C_i)+\ell_i+1+2t_i=2(p_i+t_i)+1\equiv1\pmod 4$, and
\begin{equation}
  \kappa\equiv n-\operatorname{sign}(\mathsf e)\pmod 4\;.
\end{equation}
For even $p+q$, $\sigma$ is even and $\e(-\sigma\kappa/8)$ depends only on $\kappa\bmod4$. Since $Z^{\rm can}_{\rm CS}[-M]=\overline{Z^{\rm can}_{\rm CS}[M]}$, this proves~\eqref{eq:ZMcan}.

\subsection{Restoring the \texorpdfstring{$\eta$}{eta}-factors}
\label{app:eta}
The physical partition functions are $Z_x$ in~\eqref{eq:Zx}. Let $\cO^{\rm phys}_{g,n;\chi}$ be~\eqref{eq:Odef} with $\Theta_x$ replaced by $Z_x$. Define the eta multiplier $v_\eta$ by
\begin{equation}
  \eta(\gamma\tau)=v_\eta(\gamma)\,(c\tau+d)^{1/2}\,\eta(\tau)\;,
\end{equation}
with the principal branch. For the representatives of Section~\ref{sec:review}, $v_\eta(\gamma)=\e\bigl(\Phi(\gamma)/24-1/8\bigr)$ for $c>0$, where $\Phi$ is the Rademacher function~\cite{RademacherGrosswald1972}, while $v_\eta(\mathbbm 1)=1$ for the identity coset $(c,d)=(0,1)$. Since $|v_\eta|=1$, we have $[\eta^p\bar\eta^q](\gamma\tau)=v_\eta(\gamma)^{\sigma}\,j_{p,q}(\gamma,\tau)\,[\eta^p\bar\eta^q](\tau)$. Inserting this into $\vev{Z_x(\tau)}=E_x(\tau)/[\eta^p\bar\eta^q](\tau)$ reproduces~\cite[(45),(58)]{Ashwinkumar:2021kav} for $c>0$, and gives for all cosets
\begin{equation}
  \vev{\cO^{\rm phys}_{g,n;\chi}(\btau;m_1,\dots,m_n)}
  =\sum_{\gamma_1,\dots,\gamma_n}\mathsf W(\bgamma)\prod_{i=1}^n\frac{1}{\eta(\gamma_i\tau_i)^p\,\overline{\eta(\gamma_i\tau_i)}^{\,q}}\;,
\end{equation}
with
\begin{equation}
  \mathsf W(\bgamma):=Z^{\rm pr}_{\rm CS}[g;\chi;\bgamma]\prod_{i=1}^nv_\eta(\gamma_i)^{\sigma}\;.
  \label{eq:physW}
\end{equation}
This includes the degenerate fillings, for which $v_\eta(\mathbbm 1)^\sigma=1$, so that the identity-coset term is simply $\delta_{x,0}/(\eta^p\bar\eta^q)(\tau_i)$. If all $c_i\ge1$, then~\eqref{eq:ZMcan} applies, and combining the $n$ factors $\e(-\sigma/8)$ from $v_\eta^\sigma$ with the phase in~\eqref{eq:ZMcan} gives, for even $p+q$,
\begin{equation}
  \mathsf W(\bgamma)=\overline{Z^{\rm can}_{\rm CS}\bigl[M(g;\chi;\bgamma)\bigr]}\;
  \e\Bigl(\frac{\sigma}{24}\Bigl(\sum_{i=1}^n\Phi(\gamma_i)-3\operatorname{sign}(\mathsf e)\Bigr)\Bigr)\;.
  \label{eq:physical}
\end{equation}
The factors $1/(\eta^p\bar\eta^q)(\gamma_i\tau_i)$ are the one-loop factors of the solid tori $V_i$, and each term is independent of the choice of coset representatives: under $\gamma_i\mapsto T^k\gamma_i$, $v_\eta(\gamma_i)^\sigma$ and $[\eta^p\bar\eta^q](\gamma_i\tau_i)$ both acquire the factor $\e(\sigma k/24)$, since $\Phi(T^k\gamma)=\Phi(\gamma)+k$ and $\eta(\tau+k)=\e(k/24)\eta(\tau)$. The phase in~\eqref{eq:physical}, however, depends on the representative, and we now fix it. Let $\Gamma^{(\alpha,\beta)}:=\bigl(\begin{smallmatrix}\alpha&\rho'\\ \beta&\delta'\end{smallmatrix}\bigr)\in\SL(2,\ZZ)$ be the gluing matrix of~\cite{Rozansky:1993zx}, and choose the representative
\begin{equation}
  \gamma_i:=S\,\bigl(\Gamma^{(\alpha_i,\beta_i)}\bigr)^{T}=\begin{pmatrix}-\rho'_i&-\delta'_i\\ \alpha_i&\beta_i\end{pmatrix}\;,
\end{equation}
which has lower row $(\alpha_i,\beta_i)$. With $\Phi(\gamma)=(a+d)/c-12s(d,c)$ for $c>0$, and $\rho'\equiv-\beta^{-1}\bmod\alpha$, the identity $s(-\beta^{-1},\alpha)=-s(\beta,\alpha)$ gives $\Phi(S\Gamma^{(\alpha,\beta)})=(\rho'-\beta)/\alpha+12s(\beta,\alpha)$ and $\Phi(\gamma)=(\beta-\rho')/\alpha-12s(\beta,\alpha)$, i.e.
\begin{equation}
  \Phi(\gamma_i)=-\Phi\bigl(S\Gamma^{(\alpha_i,\beta_i)}\bigr)
\end{equation}
exactly. We have also checked this numerically for $\alpha\le11$, $|\beta|\le15$. For this choice, the exponent in~\eqref{eq:physical} is exactly $-\bigl[\sum_i\Phi(S\Gamma^{(\alpha_i,\beta_i)})+3\operatorname{sign}(\mathsf e)\bigr]$. Any other representative $T^{k_i}\gamma_i$ shifts it by $\sum_ik_i$, which is compensated by the $\eta$-factors as explained above.

Let us compare this with the framing correction of~\cite[(3.48)]{Rozansky:1993zx}. With $B:=\sum_i\Phi(SM^{(p_i,q_i)})+3\operatorname{sign}(H/P)$ and the central charge $c=3(K-2)/K$~\cite[(3.47)]{Rozansky:1993zx}, that correction equals $\exp(i\pi cB/12)$. The phase in~\eqref{eq:physical} has the same form with $c$ replaced by $\sigma$, up to the overall sign, which reflects the orientation convention of Section~\ref{sec:dict}. The formula of~\cite{Rozansky:1993zx} is written for three fibers with gluing matrices $M^{(p_i,q_i)}$, which correspond to our $\Gamma^{(\alpha_i,\beta_i)}$, uses $\Phi(SM^{(p_i,q_i)})=\Phi(M^{(p_i,q_i)})-3\operatorname{sign}(p_iq_i)$~\cite[(3.46)]{Rozansky:1993zx}, and is extended to $n$ fibers in~\cite[Section~5.2]{Rozansky:1993zx}. It reduces the invariant from the framing defined by the gluing matrices to the standard framing of~\cite{Rozansky:1993zx}, whose signature phase~\cite[(2.22)]{Rozansky:1993zx} has the same form as that of~\eqref{eq:surgery}, with $c$ replaced by $\sigma$. There $H/P=\sum_iq_i/p_i$, which equals $\mathsf e$ once $\chi$ is absorbed into the Seifert invariants. The CFT therefore weights each Seifert manifold by its Chern--Simons invariant in the framing defined by the solid tori $V_i$.

\section{Non-orientable bases: derivation}
\label{app:nonor}
We derive the crosscap functional~\eqref{eq:crosscap}. The idea is to compute the functional of $X:=K\tilde\times I$ in a Seifert fibration with an orientable base, where~\eqref{eq:core} applies, and then to change to the fiber basis of the fibration over the M\"obius band.

\paragraph{Two fibrations.}
The boundary of $X$ is a single torus, and $X$ admits two Seifert fibrations~\cite[Section~5.4(iii)]{Orlik1972}: (A) over the M\"obius band, i.e.\ over $N_1$ with one boundary circle, and (B) over the disc with two exceptional fibers $(2,1)$ and $(2,-1)$. On the boundary torus, the regular fiber $\mathfrak f$ of (A) is the section boundary of (B). To see this, note that in (B) we have
\[
  H_1(X)=\langle\mathfrak l_1,\mathfrak l_2,\mathfrak m\mid2\mathfrak l_1+\mathfrak m=0,\ 2\mathfrak l_2-\mathfrak m=0\rangle\;,
\]
so that the section boundary $\mathfrak l:=-\mathfrak l_1-\mathfrak l_2$ satisfies $2\mathfrak l=0$, while $\mathfrak m$ has infinite order. In (A), the fiber is reversed along the core of the M\"obius band, so that $2\mathfrak f=0$. Since a primitive class on the boundary torus that is torsion in $H_1(X)$ is unique up to sign, $\mathfrak f=\pm\mathfrak l$.

\paragraph{The functional in fibration (B).}
Fibration (B) is the core~\eqref{eq:core} with $g=0$, $n=3$ and $\chi=0$, with two of its legs filled by the solid tori of the exceptional fibers. In the fiber basis of (B), with label $x$ the holonomy around $\mathfrak m$, the coefficient of $\bra{x}$ is therefore
\begin{equation}
  \Psi^{(\rm B)}_x\doteqf|\Ds|^{1/2}\,\varepsilon_{Q,x}(2,1)\,\varepsilon_{Q,x}(2,-1)\;.
\end{equation}
The representatives $\gamma_{2,\pm1}=ST^{\mp2}S^{-1}$ (up to the sign of $\gamma_{2,-1}$, under which $\varepsilon_{Q,x}$ is unchanged, since $\varepsilon_{Q,-x}=\varepsilon_{Q,x}$) and~\eqref{eq:weil} give
\begin{equation}
  \varepsilon_{Q,x}(2,\pm1)\doteqf\frac{1}{|\Ds|}\sum_{y\in\Ds}\e\bigl(\pm2q(y)-(x,y)\bigr)\;.
\end{equation}

\paragraph{Change to fibration (A).}
By the previous step, the fiber basis of (A), with label $w$ the holonomy around $\mathfrak f=\pm\mathfrak l$, is related to that of (B) by $\rho(S)$. The two choices of sign differ by $w\mapsto-w$, which does not affect the result below. Hence
\begin{equation}
  \begin{aligned}
    \Psi_w & \doteqf\frac{1}{\sqrt{|\Ds|}}\sum_{x\in\Ds}\e\bigl(-(x,w)\bigr)\,\Psi^{(\rm B)}_x
    =\frac{1}{|\Ds|^{2}}\sum_{x,y,y'\in\Ds}\e\bigl(2q(y)-2q(y')-(x,y+y'+w)\bigr)                    \\
           & =\frac{1}{|\Ds|}\sum_{y\in\Ds}\e\bigl(2q(y)-2q(w+y)\bigr)
    =\e\bigl(-2q(w)\bigr)\,\frac{1}{|\Ds|}\sum_{y\in\Ds}\e\bigl(-2(w,y)\bigr)
    =\e\bigl(-2q(w)\bigr)\,\delta_{2w,0}\;.
  \end{aligned}
\end{equation}
In the second line, the sum over $x$ has set $y'=-w-y$, and we have used $q(-v)=q(v)$ and $q(w+y)=q(w)+q(y)+(w,y)$. For $2w=0$ we have $4q(w)=q(2w)=0$ mod $1$, so that $\e(-2q(w))=\e(2q(w))=\nu_w$. This is~\eqref{eq:crosscap}.

The support on $\Ds[2]$ can be understood directly. Since $2\mathfrak f=0$ in $H_1(X)$, a Wilson line along $2\mathfrak f$ bounds a surface in $X$ and acts trivially on the functional. Its eigenvalue on $\bra{w}$ is $\e(\pm2(w,y))$ for charge $y$, so that $(2w,y)\in\ZZ$ for all $y$, i.e.\ $2w=0$.

\paragraph{Gluing crosscaps.}
In the two-dimensional TQFT of Section~\ref{sec:2d}, \eqref{eq:crosscap} is the crosscap state: crosscaps project onto the sectors fixed by charge conjugation $x\mapsto-x$, with weight $\nu_x\theta_x^{1/2}$. The fiber of (A) is the fiber of the core, so we can glue $X$ to a leg of the orientable core by contracting in the fiber basis. Gluing $h$ crosscaps into~\eqref{eq:core} with $g=0$ and $n+h$ legs gives
\[
  |\Ds|^{-1+\frac{n+h}{2}}\sum_{x\in\Ds[2]}\nu_x^{\,h}\,\e\bigl(-\chi q(x)\bigr)\,\bra{x}_1\otimes\cdots\otimes\bra{x}_n\;,
\]
which is~\eqref{eq:core-nonor}. This is consistent with Section~\ref{sec:selfglue}. Gluing two legs of the core~\eqref{eq:core} with the fiber reversed, through the cylinder $\sum_x\ket{x,-x}$, imposes $x=-x$ and leaves $|\Ds|^{g+n/2}\sum_{x\in\Ds[2]}$ on the remaining $n$ legs. The base is then $N_{2g+2}$, and this agrees with~\eqref{eq:core-nonor} for $h=2g+2$, since $\nu_x^2=1$. The partition functions of Seifert manifolds over $N_h$ and the replica observable follow from~\eqref{eq:ZM} and~\eqref{eq:Odef} with $\cV_{g,n;\chi}$ replaced by $\cV^{N}_{h,n;\chi}$, and the Seifert--Eisenstein series is~\eqref{eq:Z-nonor}. For $\Ds=\ZZ_3$ (Section~\ref{sec:example}), $\cZ^{N}_{h,n;\chi}=3^{h/2-1+n/2}\prod_iE_0(\tau_i)$.
\section{Orbifold geometries}
\label{sec:orbifolds}
The $\ZZ_m$ orbifolds $M_{(c,d)}/\ZZ_m$ of~\cite{Maloney:2007ud}, excluded in~\cite[fn.~9]{Ashwinkumar:2021kav}, are Seifert fibered orbifolds. The $\ZZ_m$ fixes the horizon geodesic, which becomes an orbifold locus with cone angle $2\pi/m$ and isotropy group $\ZZ_m$. This should be distinguished from an exceptional fiber of order $\alpha$ in a Seifert fibration of a smooth manifold, such as those of Section~\ref{sec:seifert}: there the total space is smooth with trivial isotropy, and $\alpha$ only records that the nearby regular fibers wind $\alpha$ times around the exceptional one. The two cases look alike only after projection to the base, where both give a cone point. The orbifolds are, however, invisible to the Abelian TQFT. The underlying topological space of $M_{(c,d)}/\ZZ_m$ is again a solid torus, whose meridian is a primitive slope on the boundary torus, so its Chern--Simons state is one of the vectors $\rho(\gamma)^{-1}\ket0$ already appearing in~\eqref{eq:E-def}. The orbifolds therefore differ from the terms of~\eqref{eq:E-def} only in their gravitational weight: the boundary torus is the quotient of the boundary of the $m$-fold cover, so the automorphy factor is evaluated at a modulus $\tilde\tau:=(a\tau+b)/d$ with $ad=m$. This explains why they are not needed to reproduce $\vev{\Theta(\tau)}$.

Such weights arise instead from Narain observables on $m$-fold covering tori. Since~\eqref{eq:SW1} holds at any modulus, $\vev{\Theta(\tilde\tau)}=E(\tilde\tau)$, whose Poincar\'e expansion runs over the products $\gamma\delta$ with $\gamma\in\Ginf\backslash\SL(2,\ZZ)$ and $\delta:=\bigl(\begin{smallmatrix}a&b\\0&d\end{smallmatrix}\bigr)$, which are integer matrices of determinant $m$. These are the solid tori $\tilde V$ bounded by the covering torus $\tilde{\mathbb T}=\CC/\tilde\Lambda_\tau$, with $\tilde\Lambda_\tau\subset\Lambda_\tau=\ZZ+\tau\ZZ$ of index $m$. The deck group $\Lambda_\tau/\tilde\Lambda_\tau$ always extends to $\tilde V$, acting by rotations of the meridian disc combined with translations along the core. Let $\tilde{\mathfrak m}\in\tilde\Lambda_\tau$ be the meridian of $\tilde V$. If $\tilde{\mathfrak m}$ is primitive in $\Lambda_\tau$, the action is free, and $\tilde V/(\Lambda_\tau/\tilde\Lambda_\tau)$ is a smooth solid torus, i.e.\ a term of~\cite{Ashwinkumar:2021kav} with a different weight. If instead $\tilde{\mathfrak m}=k\mathfrak m'$ with $\mathfrak m'\in\Lambda_\tau$ primitive and $k>1$, the elements of the subgroup
\[
  \frac{(\QQ\tilde{\mathfrak m}\cap\Lambda_\tau)+\tilde\Lambda_\tau}{\tilde\Lambda_\tau}=\frac{\ZZ\mathfrak m'+\tilde\Lambda_\tau}{\tilde\Lambda_\tau}\cong\ZZ_k
\]
of the deck group act as rotations of the meridian disc and fix the core. (Since $\tilde{\mathfrak m}$ is primitive in $\tilde\Lambda_\tau$, $\ZZ\mathfrak m'\cap\tilde\Lambda_\tau=\ZZ\tilde{\mathfrak m}$.) The quotient therefore has an orbifold core with cone angle $2\pi/k$ and isotropy $\ZZ_k$, in the sense explained above, rather than an exceptional fiber of a smooth Seifert manifold. Here $k$ divides $m$, and the deck group is the cyclic group $\ZZ_m$ fixing the core only for $k=m$. The $\ZZ_m$ orbifolds $M_{(c,d)}/\ZZ_m$ are the terms with $k=m$, i.e.\ $\Lambda_\tau/\tilde\Lambda_\tau\cong\ZZ_m$ generated by $\mathfrak m'=\tilde{\mathfrak m}/m$. The Hecke-type combination $\sum_{ad=m,\,b\bmod d}\Theta\bigl((a\tau+b)/d\bigr)$, made covariant using the vector-valued Hecke operators of~\cite{BruinierStein2010} (for $m$ coprime to the level of $\Ds$), is therefore the natural ensemble observable that includes $M_{(c,d)}/\ZZ_m$.

\bibliographystyle{ytphys}
\bibliography{draft}

\end{document}